\documentclass[conference]{IEEEtran}
\IEEEoverridecommandlockouts
\usepackage{cite}
\usepackage{amsmath,amssymb,amsfonts}
\usepackage{graphicx}
\usepackage{textcomp}
\usepackage{xcolor}
\usepackage{tikz}
\usepackage{amsmath}

\usepackage{filecontents}
\usepackage{booktabs}
\usepackage{multirow}

\usepackage{graphicx}
\usepackage{subfigure}
\usepackage{algorithm}
\usepackage{algpseudocode}
\usepackage{amsmath}
\usepackage{balance}
\usepackage{url}
\usepackage{soul}

\def\BibTeX{{\rm B\kern-.05em{\sc i\kern-.025em b}\kern-.08em
    T\kern-.1667em\lower.7ex\hbox{E}\kern-.125emX}}
\begin{document}

\title{APFuzz: Towards Automatic Greybox Protocol Fuzzing}

% \and
% \and
% \and
% \and
% \and

\author{
\IEEEauthorblockN{Yu Wang\IEEEauthorrefmark{1},
Yang Xiang\IEEEauthorrefmark{1},
Chandra Thapa\IEEEauthorrefmark{2},
Hajime Suzuki\IEEEauthorrefmark{2}}
\IEEEauthorblockA{\IEEEauthorrefmark{1}\textit{Swinburne University of Technology}, Australia\\
\{hf98403@gmail.com, yxiang@swin.edu.au\}}
\IEEEauthorblockA{\IEEEauthorrefmark{2}\textit{CSIRO Data61}, Australia\\
\{chandra.thapa@data61.csiro.au, Hajime.Suzuki@data61.csiro.au\}}
}
\maketitle

\begin{abstract}
%-------------------------------------------------------------------------------
%As 6G networks are expected to support a wide array of new applications and services, ensuring the security of communication protocols is crucial.
Greybox protocol fuzzing is a random testing approach for stateful protocol implementations, where the input is protocol messages generated from mutations of seeds, and the search in the input space is driven by the feedback on coverage of both code and state.
State model and message model are the core components of communication protocols, which also have significant impacts on protocol fuzzing.
In this work, we propose APFuzz (Automatic greybox Protocol Fuzzer) with novel designs to increase the smartness of greybox protocol fuzzers from the perspectives of both the state model and the message model.
On the one hand, APFuzz employs a two-stage process of static and dynamic analysis to automatically identify state variables, which are then used to infer an accurate state model during fuzzing. 
On the other hand, APFuzz introduces field-level mutation operations for binary protocols, leveraging message structure awareness enabled by Large Language Models.
%State model is the basis of state-coverage based fuzzing, where the key challenge is how to identify the state of the protocol during fuzzing.
%However, different state definitions and representations are used in existing greybox protocol fuzzers. 
%It is difficult to compare them for providing insight on their internal properties and effectiveness on driving the protocol fuzzing campaign. 
%From the perspective of message model, recent studies show that Lager Language Models are capable of providing the knowledge of open protocols without requiring much manual efforts. 
%They have been successfully applied to improve the input structure awareness for text-based protocols, but the feasibility on binary protocols remains unexplored. 
We conduct extensive experiments on a public protocol fuzzing benchmark, comparing APFuzz with the baseline fuzzer AFLNET as well as several state-of-the-art greybox protocol fuzzers.
\end{abstract}


\section{Introduction}
%-------------------------------------------------------------------------------

%As 6G networks are expected to support a wide array of new applications and services, ensuring the security of communication protocols is crucial. Greybox protocol fuzzing helps identify vulnerabilities in protocol implementations, thereby enhancing the overall security posture of 6G networks.

% Fuzzing
The concept of fuzzing involves testing a program by providing a large amount of random data as input to uncover bugs and vulnerabilities \cite{Zhu2022Roadmap}. Originally, the word fuzz was used to name the random character generator in the UNIX utility testing tools built by Miller et al. \cite{Miller1990Fuzzing}. The technique has evolved since then, and it is behind the discovery of many serious security vulnerabilities in lots of popular software and critical systems, including communication protocols \cite{Bohme2021Challenges}.

% Fuzzing stateful
Most fuzzing tools are designed for testing stateless programs. There are additional challenges when it comes to testing stateful programs \cite{Daniele2024Survey}. On one hand, the same input can result in different reactions of a stateful program depending on its current state, which is determined by previous inputs. On the other hand, the stateful bugs in these programs can only be triggered in certain states hidden in the deep state space \cite{Aschermann2020Ijon}. To expose such a stateful bug, the fuzzer needs to first send a specific series of inputs to reach the correct state and then send the fuzzing inputs to trigger the bug.

% Fuzzing protocol
A typical type of stateful program is protocol implementation. In general, a communication protocol defines a set of rules that allow two or more entities to exchange various messages in a conversation. For example, syntax rules specify the progress of the conversation, which gives the context for evaluating each received message, while semantics rules specify the meaning of data, which describes the format to interpret each received message. In other words, protocol syntax defines the state model, and protocol semantics defines the message model. Traditional protocol fuzzing tools (e.g., SNOOZE \cite{Banks2006SNOOZE}, Sulley, BooFuzz from academia, and Peach, beSTORM from industry) are based on the blackbox approach. That is, the fuzzers are provided with the state model and message model of the target protocol, and they traverse the state model to send input test cases generated based on the message model. The effectiveness of blackbox protocol fuzzing depends on the quality of the state model and message model, which are usually manually crafted based on protocol specifications and can be error-prone.

% Greybox fuzzing
Coverage-based greybox fuzzing \cite{Gao2023Plateau} uses lightweight code instrumentation to obtain coverage information to guide the input space exploration. For example, American Fuzzy Loop (AFL) \cite{AFL,Fioraldi2020AFL++}, one of the most widely-used coverage-based greybox fuzzers, uses the information about block transitions and branch hit counts to guide input mutations. To enable greybox protocol fuzzing, AFLNET \cite{Pham2020AFLNET} extends AFL by introducing the concept of state coverage and proposing a state-oriented scheduling scheme. Since then, a series of variants of AFLNET have been proposed, which focus on different aspects of improvements. However, there are still some open problems in greybox protocol fuzzing.

% KEY CHALLENGE: PROTOCOL STATE MODEL
\textit{State representation.}
The feedback of state coverage is the core of stateful greybox fuzzing. However, recognizing the protocol states during fuzzing is not straightforward. Existing fuzzers adopt different state definitions and representations. For example, AFLNET uses the status codes in server responses to represent protocol states, while other fuzzers adopt variable-based state representation (e.g., NSFuzz \cite{Qin2023NSFuzz} and SGFuzz \cite{Ba2022SGFuzz}) and long-lived memory data based fuzzy state representation \cite{Natella2022StateAFL}. Moreover, most state representation schemes implicitly require human intervention. For example, given a new target protocol to test, AFLNET requires users to write the code to interpret server responses and extract status codes; NSFuzz requires users to read the source code and use its annotation API to manually specify the correct state variables (as the heuristic-based identification brings lots of false positives and false negatives). Therefore, it is still challenging to learn the state representation effectively and automatically. 
%The relations of these different state representations and their impacts on coverage-based protocol fuzzing have not been studied systematically. 

% KEY CHALLENGE: PROTOCOL MESSAGE MODEL
\textit{Input structure awareness.}
AFLNET introduces message-level mutating operations on top of the byte-level mutations implemented in AFL. However, random mutation is likely to break the message format and get it rejected by the server. A recent study \cite{Meng2024ChatAFL} proposes to use a Large Language Model (LLM) to extract grammar patterns for text-based protocols, which can be applied to the messages to match the keyword fields and mark the data fields as mutable regions. This realizes structure-aware field-based mutations. However, this approach does not support binary protocols.

In this paper, we propose APFuzz, a coverage-based greybox protocol fuzzing tool. 
APFuzz has two key features.
On the one hand, it introduces state representation learning, enabling the automatic identification of state variables in source code and efficient state tracking during fuzzing.
The learning process consists of two stages.
The first stage involves using static analysis to identify potential state variables and the second stage focuses on filtering out the noises by using a lightweight dynamic analysis approach. 
On the other hand, it leverages LLM for input structure learning, enabling field-based mutation operators for binary protocols.
Specifically, the LLM is used to interpret the message structure of the initial seeds and derive knowledge about the fields (e.g., field type, bit range, value range). 
Such message structure information is valuable for mutations to generate valid messages to reach deep-state space. 
We conduct extensive experiments based on ProFuzzBench \cite{Natella2021ProFuzzBench}, the benchmark for protocol fuzzing. We compare APFuzz with the baseline fuzzer AFLNET as well as a number of state-of-the-art greybox protocol fuzzers. 

% CONTRIBUTIONS
In summary, we make the following contributions to greybox protocol fuzzing:
\begin{itemize}
    %\item[$\bullet$] We investigate the influence of state representations in greybox protocol fuzzing. In particular, we integrate three commonly used state representation approaches (i.e., response-based, memory-based and variable-based) and present a comprehensive analysis on their internal properties and their impacts on coverage-based fuzzing.
    \item[$\bullet$] We propose a state representation learning method that automatically identifies the state variables in the source code of the target server. This enables accurate and efficient protocol state tracking and state model inference.
    \item[$\bullet$] We introduce field-based mutations through an input structure learning method that leverages LLM to acquire message structure knowledge. This helps generate valid test cases and explore the deep state space of protocols.
    \item[$\bullet$] We conduct and present a comprehensive evaluation of our framework against the baseline and the state-of-the-art greybox protocol fuzzers based on the widely used protocol fuzzing benchmark ProFuzzBench.
\end{itemize}

\section{Background and Motivations}
%-------------------------------------------------------------------------------

Recently, the problem of protocol fuzzing has attracted a lot of research interest. Daniele et al. \cite{Daniele2024Survey} present a comprehensive survey of fuzzers for stateful systems, and Zhang et al. \cite{Zhang2023Survey} present a survey on the development of fuzzers for network protocols. In this section, we start by briefly introducing the latest progress of coverage-based greybox protocol fuzzing. Then, we discuss the key open issues and our motivations.

\subsection{Greybox Protocol Fuzzing}
We take AFLNET as the baseline greybox protocol fuzzer to facilitate the discussion on the recent progress in this area.

AFLNET, proposed by Pham et al. \cite{Pham2020AFLNET} in 2020, is the first greybox fuzzer for protocol implementations. It is implemented based on AFL \cite{AFL}, with several critical extensions to address the challenges of applying the stateless fuzzer to stateful servers. First, AFLNET defines the input as a sequence of messages that the client sends to the server in a conversation (i.e., the client requests). Second, AFLNET seeks to maximize not only the code coverage but also the state coverage. To do this, it observes the server state after sending each message in order to associate a state sequence with the input message sequence. Then, an input is considered as interesting (i.e., to be added to the seed corpus) if any new state or state transition is found in its associated state sequence, as well as if any new code branch is executed in the server. AFLNET also introduces a state-oriented scheduling scheme and uses some heuristics to prioritize the rarely executed states as well as the states that are successful in increasing code or state coverage when selected previously. Third, AFLNET supports message-level mutation operators in addition to the byte-level ones. Finally, AFLNET makes some technical designs to enable server fuzzing, such as enabling sockets for network I/O, adding delays between requests for fuzzer-server synchronization, and resetting the server after executing each input.

The speed of protocol fuzzing can be measured by the throughput in terms of messages per second or executions per second. AFLNET is significantly slower than AFL because, for each input test case, AFLNET goes through the process of initializing the server, sending messages one after one, and waiting in between, cleaning up and terminating the server. Fu et al. \cite{Fu2023HNPFuzzer} analyze the time consumption by dividing this process into four stages, including initialization, socket-based inter-process communication (IPC), synchronization, and other behaviors (e.g., termination). They find that socket-based IPC consumes the majority of time for most target servers in ProFuzzBench \cite{Natella2021ProFuzzBench}, while synchronization and initialization also take significant amounts of time for several servers. Thus, they extend AFLNET with shared-memory-based IPC and synchronization for message exchanges, and they also introduce the persistent mode to avoid server resets whenever possible. The resulting tool, HNPFuzzer, is able to achieve much higher fuzzing throughput than AFLNET. There are other approaches to speed up the process. For example, SNPSFuzzer \cite{Li2022SNPSFuzzer} takes process snapshots by using the CRIU tool (Checkpoint/Restore in User Space) to skip the time of initializing the server and sending the prefix messages before the mutated ones. Similarly, Nyx-net \cite{Schumilo2022Nyx-net} adopts hypervisor-based snapshots to allow fast server resets and emulates the APIs for networking I/O to speed up message exchanges. SnapFuzz \cite{Andronidis2022SnapFuzz} intercepts the communication between the target server and the fuzzing environment via binary rewriting in order to eliminate communication and synchronization delays as well as to improve the speed of file access, clean-ups, and state resets.

The smartness of greybox protocol fuzzing is also important, which allows the fuzzer to cover more code with fewer tests. Recent studies in this direction mainly involve investigations of different state representations and input structure awareness. In the following sections, we discuss these two aspects. 

\subsection{Protocol State Awareness}

Identifying server or protocol states during fuzzing is not straightforward. AFLNET infers state information by extracting the status code in server responses. However, the response codes are not available in all protocols, and by design, they are not directly linked to protocol states (e.g., HTTP and FTP use the code 200 to acknowledge the successful completion of different kinds of requests and commands). Besides, manual efforts and protocol knowledge are required to tell the fuzzer how to extract the response codes. To overcome these limitations, several different solutions are proposed in recent studies. 

Natella \cite{Natella2022StateAFL} proposes StateAFL that uses the data in long-lived memory areas to infer the server state. In essence, StateAFL learns the fuzzy representations of in-memory server states. This is done in a fully automated manner, though it does introduce computational overhead to some extent. 

Another solution is to identify the state variables in the source code, whose values can be considered as the ground truth of state identification because they are used internally by the servers to maintain the states of conversations. Ba et al. \cite{Ba2022SGFuzz} propose SGFuzz, which monitors changes in the values of all variables of enumerated type and gradually constructs a state transition tree during fuzzing. There are some false positives as the enumerated type variables can be used not only for state representations but also for error codes, response codes, configuration settings, and so on. SGFuzz mitigates the effects of such noises by using tree truncating and energy scheduling schemes. Qin et al. \cite{Qin2023NSFuzz} propose NSFuzz, which uses static analysis and a heuristics-based approach to locate the network event loop that handles client requests and identify state variables within the loop. NSFuzz also allows users to manually specify the state variables through its annotation APIs.

% State-coverage optimization is the core of stateful greybox fuzzing. However, the definitions of state in existing studies are inconsistent. For example, AFLNET uses server response codes as state representation, StateAFL uses the fuzzy hash of data in the long-lived memory areas as state representation, while NSFuzz and SGFuzz adopt variable-based state representation. This inconsistency leads to completely different designs on various state-related operations, such as state identification, state model inference, and state coverage optimization. Therefore, it is difficult to compare different state definitions and get insights into their internal properties and relations. 

\textbf{Motivation:} State-coverage optimization is the core of stateful greybox fuzzing. However, the definitions of state in existing studies are inconsistent, and it leads to different designs on various state-related operations, such as state identification, state model inference, and state coverage optimization. In addition, most fuzzers require manual efforts to extract state information. For example, given a new target to test, the users need to know about the server response format and provide AFLNET with a script for extracting response codes or need to inspect the source code and identify the accurate state variables for NSFuzz. Driven by these limitations and challenges, this work aims to explore accurate and automated state representation learning.

\subsection{Message Structure Awareness}

Awareness of input structure can be beneficial to coverage-based greybox fuzzing \cite{Pham2021Smart}. In protocol fuzzing, the inputs are message sequences, where the order of messages is governed by protocol syntax, and the format of messages is defined by protocol semantics. AFLNET supports two types of mutations. The first type is the byte-level mutations inherited from AFL, such as random bit-flip, byte-insert, byte-delete, byte-clone, and so on \cite{chen2018systematic}. These mutations are likely to challenge the semantics rules and exercise the message interpreters in protocol implementations. The other type is message-level mutations, including random message insert, delete, replace, and duplicate. These mutations are likely to challenge the syntax rules and exercise the state model of the servers. 

Fuzzing protocols by using random mutations and simple message-level mutations is lightweight, as they do not rely on complex computations or knowledge about the target \cite{chipounov2011s2e, nethercote2007valgrind}. However, these mutations are inefficient for exploring deep-state space and uncovering deep-state bugs, as most of such arbitrary mutations are likely to be invalid and break the conversations in the early stage. Traditional protocol fuzzers address this limitation by applying grammar-based mutations using predefined rules provided by users \cite{Banks2006SNOOZE, kaksonen2001software, boofuzz}. While effective, these methods are both labor-intensive and time-consuming.

The recent advances in Large Language Models (LLMs) allow fuzzers to incorporate protocol grammar for mutations without human input. For example, Meng et al. \cite{Meng2024ChatAFL} propose to extract message grammar patterns from ChatGPT to enable fine-grained structure-aware mutations for text-based protocols in ChatAFL. Given a message to mutate, ChatAFL first applies the grammar patterns to identify the message type. It then matches the header fields in the message and marks the data fields as mutable regions. Besides, ChatAFL also uses the LLM to generate seeds for increasing the diversity of initial seeds and for surpassing the coverage plateau during fuzzing. Although the performance is promising, ChatAFL works only on text-based protocols by design. It is not applicable to binary protocols, which use well-defined field structures instead of text-based commands and keywords.

\textbf{Motivation:} Binary protocols are widely used in practice, especially for communication systems like 5G/6G. Motivated by the success of LLMs in deriving text-based protocol patterns, this work aims to extend the grammar-guided mutations to binary protocols. In particular, we are interested in the field-level grammar that specifies bit ranges within a single message and serves as field-level mutation guidance. Field-based mutations generate more valid test cases, thus have the potentials to deliver more efficient exploration in the input space as well as the deep-state space.

\section{Methodology}
\label{sec:methodology}
%-------------------------------------------------------------------------------

% APFuzz is a coverage-based greybox fuzzing tool that aims to be "smart" in terms of being highly effectiveness while not requiring any human input. 

This section presents APFuzz, the proposed coverage-based greybox protocol fuzzing tool. 
First, we introduce the general framework and discuss the assumptions and the basic building blocks. 
Then, we describe the key features of APFuzz in detail, including proposed methods for state representation learning and input structure learning. 
%Next, we describe the key features in details, including a novel method to enable LLM-based protocol awareness for binary protocols, the supported schemes for state variable identification, and the scheduling and mutation strategies. 
%Finally, we summarize with the implementation of the fuzzing tool.

\subsection{Framework}

\begin{algorithm}[h]
  \caption{APFuzz}
  \label{algo:APFuzz}
  \renewcommand{\algorithmicrequire}{ \textbf{Input:}}
  \renewcommand{\algorithmicensure}{ \textbf{Output:}}
  \begin{algorithmic}[1]
  \Require $P_0$: protocol implementation
  \Require $p$: protocol name
  \Require $C$: initial seed corpus
  \Require $T$: total fuzzing time
  \Ensure $C$: final seed queue
  \Ensure $C_X$: crashing seeds
  \State $CandidateVariables \gets$ \textsc{StaticAnalysis}
  \State $P_s \gets$ \textsc{InstrumentState} ($P_0$)
  \State $SV \gets$ \textsc{DynamicAnalysis}($P_s, CandidateVariables$)
  \State $P_f \gets$ \textsc{InstrumentFuzz} ($P_0$)
  \State $G \gets$ \textsc{ChatLLM} ($p$, $C$)
  \State $S \gets \emptyset$
  \Repeat
      \State $s \gets$ \textsc{ChooseState} ($S$)
      \State $Messages~M$ $ \gets$ \textsc{ChooseSequence} ($C$, $s$)
      \State $\langle M_1, M_2, M_3 \rangle \gets M$
      \For {$i$ from 1 to \textsc{AssignEnergy}($M$)}
              \If {\textsc{Random}() $< \epsilon$}
                  \State $M_2' \gets$ \textsc{GrammarBasedMutate}($M_2$, $G$)
                  \State $M' \gets \langle M_1, M_2', M_3 \rangle$
              \Else
                  \State $M' \gets \langle M_1, \textsc{RandMutate}(M_2), M_3 \rangle$
              \EndIf
          \State $s' \gets$ \textsc{SendToServer}($P_f$, $M'$, $SV$)
          \State $S \gets$ \textsc{UpdateStateMachine}($s', S$) 
          \If {\textsc{IsCrashes}($M'$, $P_f$)}
              \State $C_X \gets C_X \cup \{M'\}$
          \ElsIf {\textsc{IsInteresting}($M'$, $P_f$, $S$)}
              \State $C \gets C \cup \{(M', R')\}$
          \EndIf
      \EndFor
  \Until timeout $T$ reached or abort-signal
  \end{algorithmic}
\end{algorithm}

\begin{figure}
    \centering
    \includegraphics[width=1\linewidth]{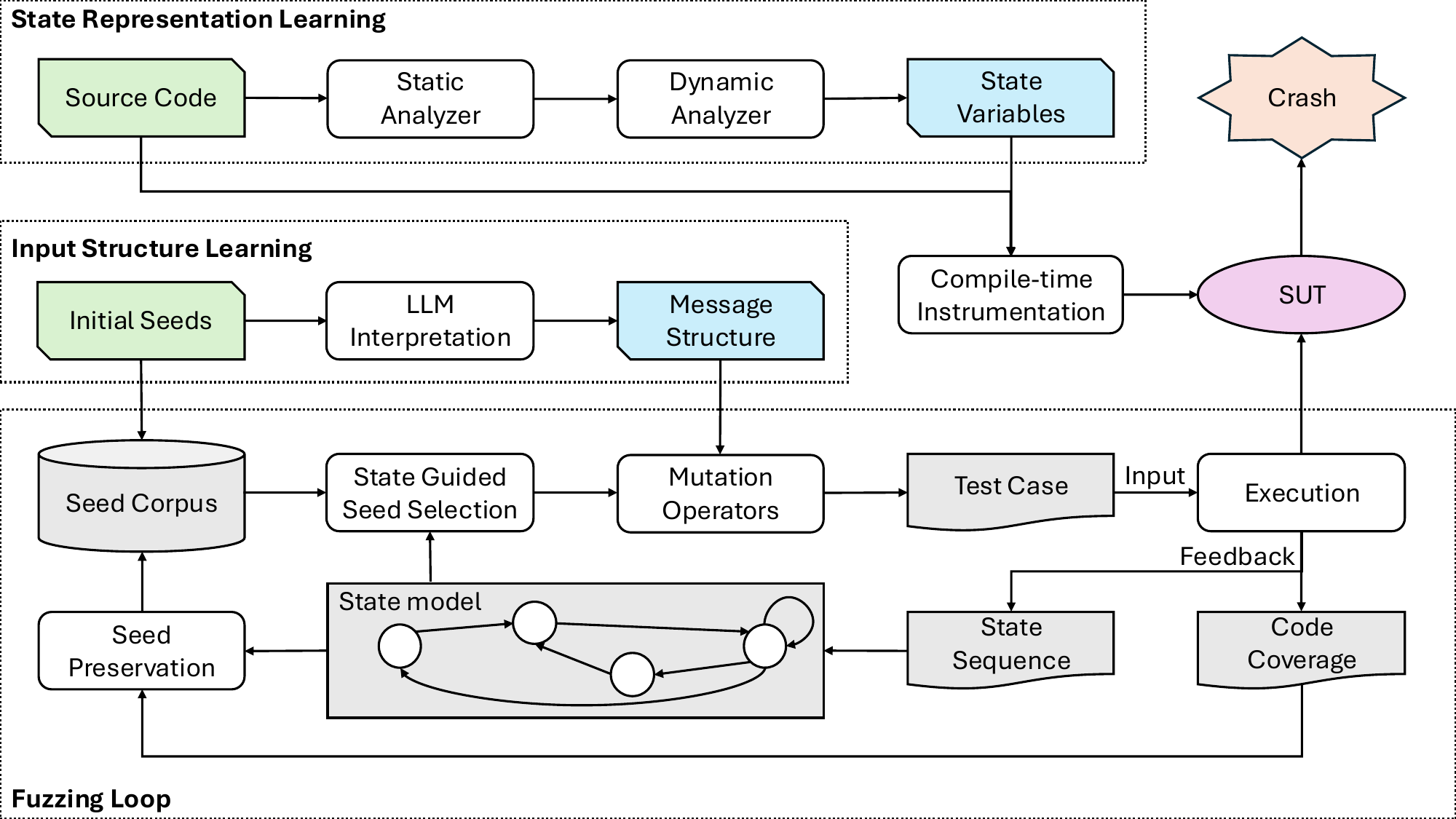}
    \caption{System framework}
    \label{fig_framework}
\end{figure}

The basic logic of greybox protocol fuzzing is the loop of automated state model inference and coverage-based fuzzing. On the one hand, the fuzzer generates new inputs to explore the unknown state space of the SUT (Server Under Test), with the goal of maximizing coverage in the state model. On the other hand, it takes feedback from the SUT to infer the states reached by each input and update the state model if any new state or state transition is found. APFuzz follows the same logic. It incrementally learns a state model of the SUT from scratch, which is used to guide the fuzzing campaign in traversing deep state space. The overall framework of APFuzz is illustrated in Fig. \ref{fig_framework}. 

\subsubsection{Assumptions} There are some prerequisites in coverage-based greybox protocol fuzzing. We first discuss the assumptions and clarify the scope of our work.

\textit{Source Code.} Like other greybox fuzzers, APFuzz requires access to the source code of the target server. It is needed for applying compile-time instrumentation to inject additional instructions into the server during compilation to collect feedback about state information, code coverage, and loop synchronization in the runtime.

\textit{Protocol Specification.}
Greybox protocol fuzzing does not require knowledge about the target protocols. ChatAFL is an exception \cite{Meng2024ChatAFL}. Although ChatAFL does not explicitly require access to protocol specifications, it does assume that ChatGPT knows about the target protocols (e.g., protocol specification documents might be included in ChatGPT's training data). Specifically, it uses ChatGPT as an oracle to derive message patterns and generate new seeds with unseen message types or unknown state transition paths for text-based protocols. APFuzz follows the same assumption, and we show that LLMs can be used to interpret not only text-based protocols but also binary protocols. 

\textit{Roles.} Existing protocol fuzzing research mainly focuses on the client/server paradigm \cite{Daniele2024Survey}. That is, the target is a server, and the fuzzer takes the role of a client. We follow the same paradigm and leave the protocols involving multiple roles (e.g., the multi-party protocols) out of scope.

\subsubsection{Input}

Input is the key to fuzzing. In stateful server fuzzing, the input is a message or a sequence of messages the client sends to the server during a session. Such messages are also referred to as requests, while the messages sent by the server are responses. %During transmission, an application-layer message may be split across multiple segments in lower-level protocols, such as packets in the network layer.

\textit{Initial Seeds.} As a mutation-based approach, greybox protocol fuzzers generate new test cases by mutating the seeds. Valid seeds can serve as a good starting point for fuzzing. APFuzz follows the common practice of using some valid samples of message sequences extracted from real traffic data of the target server to form the initial seed corpus.

\textit{Test Cases.} APFuzz adopts the mutation-based approach, where a new input is generated by mutating a seed. In addition to the byte-level and message-level mutation operators used in existing fuzzers, we introduce fine-grained mutation at the level of fields for binary protocols. This is enabled by LLM-based message structure learning. More details are presented in Section \ref{sec:mutation}. 

\subsubsection{Feedback}
Coverage-based greybox fuzzing is based on feedback. For protocol fuzzers, there are two kinds of feedback, i.e., code coverage and state information. APFuzz applies lightweight compile-time instrumentation to inject additional instructions for collecting feedback to the target servers. The feedback is used for both code coverage and state coverage.

\textit{Code Coverage.}
APFuzz adopts the instrumentation of AFL to collect code coverage information. Specifically, it contains information about block transitions and branch hit counts to guide input mutations.

\textit{State Sequence.}
APFuzz collects state feedback based on state variables, i.e., the variables that the developers use to represent protocol states in their source code. We propose a state representation learning method to identify such variables in source code in a fully automatic manner. In compile-time instrumentation, we also inject instructions for tracking these state variables to obtain the state sequence in the fuzzing loop. More details are presented in Section \ref{sec:state}. 

\subsubsection{Fuzzing Loop}

The fuzzing process is iterative. In each iteration, the fuzzer selects a seed from the seed corpus based on the state model. We follow the same strategy used in AFLNET for this step. Next, the fuzzer applies mutation operators on the seed randomly to generate a new test case. The test case is then sent to the SUT for execution. This involves a sub-loop, i.e., a message-sending loop, as the test case is a sequence of messages. In this sub-loop, the fuzzer sends each message and waits for the synchronization signal, which tells that the SUT has finished processing the current message and is ready for the next. Meanwhile, the feedback of code coverage and state sequence is collected. Finally, the state model is updated based on the state sequence. If a new state, state transition, or increase in code coverage is found, the test case is considered interesting and preserved in the seed corpus. Otherwise, the test case is discarded. The overall procedure of APFuzz is given in Algorithm \ref{algo:APFuzz}.

%\subsubsection{Fuzzer} Overview of fuzzer design.
%\textit{1) The Fuzzing Loop.} % describe an iteration
%\textit{2) State Model Inference.} 
%\textit{3) Coverage-based Fuzzing.} 

\subsection{State Representation Learning}
%\subsection{Protocol State Awareness}
\label{sec:state}

%AFLNET seeks to maximize not only the code coverage but also the state coverage. To do this, it observes the server state after sending each message in order to associate a state sequence with each input message sequence. Then, an input is considered as interesting (i.e., to be added to the seed corpus) if any new state or state transition is found in its associated state sequence, as well as if any new code branch is executed in the server. AFLNET also introduces a state-oriented scheduling scheme and uses some heuristics to prioritize the rarely executed states as well as the states that are successful in increasing code or state coverage when selected previously.

%To infer the in-memory state, StateAFL instruments the server in compile-time to add instructions to probe memory allocations and network I/O operations. During fuzzing, StateAFL takes a snapshot of the long-live memory areas in each round of message exchange in a session, and it computes a fuzzy hash of the snapshots in the end of the session. The hash value is used to represent the in-memory state, and a new state is identified if the distance of the hash value and the previous most similar hash value is larger than a threshold $\epsilon$.

To infer the protocol state, APFuzz instruments the server in compile-time to add instructions to probe the values of state variables during fuzzing. In this work, we propose an automatic method to identify state variables in the source code, which first uses static analysis to identify the potential variables and then uses dynamic analysis to track these variables and filter out the noises.

%\subsubsection{State Representation Learning}

%In protocol implementations, it is common to keep track of protocol states using dedicated state variables \cite{Ba2022SGFuzz}. SGFuzz \cite{Ba2022SGFuzz} highlights a crucial aspect of identifying protocol state variables, noting that most of these variables are either assigned to or compared with named constants, which are typically encoded as enums. Since protocol states are predefined and finite, they are often represented by intuitive named constants (e.g., INIT, READY) defined in the enumerated types directives in C/C++. Consequently, SGFuzz focuses on enums to recognize state variables. Moreover, NSFuzz \cite{Qin2023NSFuzz} considers state variables to be those that are read from and written to within network event loops for the purpose of checking and updating states. Based on these assumptions, previous studies propose heuristic-based methods to extract state variables by performing static analysis or regular expression matching. However, these methods are prone to false positives (i.e., identifying variables that are not related to protocol state) and false negatives (i.e., missing actual state variables). To address this challenge, APFuzz adopts a two-stage approach to identify state variables accurately and automatically. 

In protocol implementations, it is common to keep track of protocol states using dedicated state variables \cite{Ba2022SGFuzz}. Besides, since the states in a protocol are predefined and finite, they are typically encoded using intuitively named constants (e.g., INIT, READY) that are defined in the enumerated types or \#define directives in C/C++. Moreover, state variables should be read and written within network event loops for checking and updating states \cite{Qin2023NSFuzz}. Based on these assumptions, previous studies propose heuristic-based methods to extract state variables by performing static analysis or regular expression matching. However, these methods are prone to false positives (i.e., identifying variables unrelated to protocol state) and false negatives (i.e., missing actual state variables). To address this challenge, APFuzz adopts a two-stage approach to identify state variables accurately and automatically. 

%\textit{1) Static Analysis.} The static analysis is designed to identify a collection of potential state variables from the source code. The goal in this stage is to minimize false negatives, that is not to miss any actual state variables. Therefore, we adopt the following heuristic rules. When reproducing the SGFuzz static analysis, we found that it was incomplete in identifying state variables. It only detected variables defined explicitly as enumerated types, overlooking the possibility that state variables could be of integer type but assigned enumerated values. To address this, we enhanced the regular matching to recognize integer variables assigned with enumerated values. In this paper, we refer to this improved static analysis method as SGFuzz*. Then we combine the matching rule from both NSFuzz and SGFuzz*. Therefore, we are able to obtain a set of candidate state variables with a reasonable degree of reliability. 

\textit{1) Static Analysis.} The static analysis is designed to identify a collection of potential state variables from the source code. The goal in this stage is to minimize false negatives, that is not to miss any actual state variables. Therefore, we follow the heuristic approach used in existing fuzzers. On the one hand, we include the state variables identified by NSFuzz \cite{Qin2023NSFuzz}, which are global integer variables or user-defined structure members that are both loaded and stored (with constant values) within the identified network event loops. On the other hand, we adopt the rule used in SGFuzz \cite{Ba2022SGFuzz}, which extracts the variables of the enumerated type that are assigned at least once. We extend this rule to include variables of other types, such as integers, that are assigned with enumerated type values or named constants.

% todo：conclusion

\textit{2) Dynamic Analysis.} To filter out the noises (i.e., false positives) in the results of static analysis, APFuzz utilizes a dynamic analysis technique by leveraging the LLVM compiler infrastructure to instrument the source code, allowing for runtime tracking of the variables derived in static analysis. This approach enables detailed monitoring of how these variables are read and written during execution. APFuzz employs the following rules to filter the set of state variables more accurately through the dynamic analysis: a) The number of unique values of a state variable should be no less than $min$ and less than $max$. 
%b) The maximum value for each state variable remains below a specified threshold $t$. 
b) The hit count for each value of the state variable exceeds a certain threshold $t$. In this way, APFuzz can filter out a set of program variables that are closely related to the protocol state, which further ensures the reliability of the protocol state machine inference.

%\subsubsection{State Model Inference}

\subsection{Input Structure Learning}
%\subsection{Message Structure Awareness}
\label{sec:mutation}

Understanding the target protocol is beneficial for fuzzing. For instance, blackbox protocol fuzzing generates inputs based on the protocol state model and message model. The key challenge is to obtain accurate protocol knowledge while minimizing manual efforts. Recent advancements have demonstrated that Large Language Models (LLMs) can assist fuzzers by providing a better understanding of the target protocols in an automatic manner. For example, ChatAFL \cite{Meng2024ChatAFL} uses ChatGPT to generate message patterns for text-based protocols, which specify keywords (e.g., protocol commands) and mutable ranges (e.g., data) in messages. In this section, we extend the use of LLMs to interpret the format of the messages in the seeds for binary protocols.

\subsubsection{LLM-based Seed Interpretation}

We focus on identifying the fields in the binary protocol messages and enabling field-based mutations during fuzzing. First, the messages are converted from raw format into binary text strings before feeding into the LLM conversation. Next, we design a prompt using in-context few-shot learning \cite{brown2020language} to interact with the LLM and obtain the well-formatted structure of each part or layer of the given message. The underlying assumption is that LLMs are capable of providing useful knowledge about message structures for protocols with open RFC, which is shown in previous studies \cite{Meng2024ChatAFL}. The learned field structure of message has three parts: field name, bit start, and bit length. The field name indicates the function or type of the field. The bit start specifies the starting position in the binary message string, and the bit length indicates the number of bits in the field.

\subsubsection{Mutation Scheme}

% This section introduces a mutation strategy based on field-level information extracted by the LLM. The field structure has three parts: field name, bit start, and bit length. The field name indicates the function or type of the field. The bit start specifies the starting position in the binary message string, and the bit length indicates the number of bits in the field. 

The fuzzer selects the message sequence  $M_2$  for mutation as part of the algorithm design. Field-based mutations are guided by the bit range. The fuzzer randomly selects the start bit and corresponding length within a single message chosen from the  $M_2$  sequence. The range of the chosen field is denoted as $[start\_bit, start\_bit + length]$. We then design several bit-level structure-aware mutation strategies.
%, all shown in fig. \ref{}
These strategies exhibit finer granularity compared to the message-level mutation strategy implemented in AFLNET and the byte-level structure-aware mutations in ChatAFL. Consequently, field-based mutations can impact not only text-based protocols but also binary-based protocols such as SSH and DNS.
We introduce a random value $\epsilon$, which is selected from a range of 0 to 1. When $\epsilon > 0.5$, the fuzzer utilizes the field-based mutation strategy; otherwise, it uses the classical mutations implemented in AFLNET.

% Although conducting field-based mutations can effectively generate valid messages for exploring state space quickly, to retain the fuzzer's capability of exploring corner paths, we continue to employ structure-unaware mutations. We introduce a random value $\epsilon$, which is selected from a range of 0 to 1. When $\epsilon > 0.5$, the fuzzer utilizes the field-based mutation strategy; otherwise, it uses the classical mutations implemented in AFLNET.

%\subsubsection{Mutation Operators}

The mutation scheme in APFuzz encompasses byte-level and message-level operators. The byte-level operators are derived from AFL, while the message-level operators are derived from AFLNET. Additionally, we propose field-based mutation operators in the stack stage. The mutation operators include: 1) Deterministic Operators, such as bit or byte flips, increment or decrement existing integer values in the input file by a step of one byte, increment or decrement existing integer values in the input file by a step of one byte, select one byte from a hardcoded set of integers and overwrite existing data in the input file; 2) Classical Stack Operators, such as bit or byte flips, set "interesting" bytes and words, add or subtract small integers to bytes and words, block deletion, duplication, memset, and insertion, message replacement, insertion, and duplication; 3) Field-Based Stack Operators, such as bit flips on the field range, swap bits within the field range, overwrite the field with a random value from field dictionary. 

Deterministic mutations systematically alter all bits, bytes, words, and double words in the original input. For instance, the fuzzer sequentially iterates over all bits in the original input, generating distinct fuzz inputs by inverting each bit. Subsequently, the fuzzer applies the stack stage, where multiple mutation types are applied to the same fuzz input. If $\epsilon$>0.5, the fuzzer utilizes field-based stack operators; otherwise, it employs classical stack operators.

%\subsection{The Fuzzing Loop}

%\subsubsection{Scheduling Scheme}

%\subsection{Implementation}
%Subsection text here.

%We have built a prototype of NSFuzz. The implementation of NSFuzz is about 4.5 k lines of C/C++ code and about 100 lines of Python script. In detail, we implement the static analyzer, annotation parsing engine, and compile-time instrumentation based on LLVM [11] framework, and the fuzzer engine is implemented based on AFLNET (revision 0f51f9e from January 2021).

\begin{table}
  \caption{Benchmark targets}
  \label{table_benchmark}
  \footnotesize
  \begin{center}
      \begin{tabular}{cccccccccc}
      \toprule
      \textbf{Target} & \textbf{Protocol} & \textbf{Type} &\textbf{Version}& \textbf{Seeds} &\textbf{Lines}\\
      \midrule
      Bftpd & FTP & Text & v5.7& 54 & 4.7k\\
      Exim & SMTP & Text &38903fb & 9 & 101.7k\\
      Forked-daapd & DAAP & Text & 2ca10d9 & 65 &115.9k \\
      LightFTP & FTP & Text & 5980ea1  & 10 & 4.4k\\
      Live555 & RTSP & Text & ceeb4f4 & 33 & 52.2k\\
      ProFTPD & FTP & Text & 4017eff8 & 54 & 227.7k\\
      Pure-FTPd & FTP & Text & c21b45f & 54 & 30k\\
      Kamailio & SIP & Text & 2648eb3 & 3 & 766.7k\\
      Dcmtk & DICOM & Binary & 7f8564c & 4 & 575.2k\\
      OpenSSH & SSH & Binary & 7cfea58 & 22 & 97.3k\\
      OpenSSL & TLS & Binary & 437435 & 8 & 441.8k\\
      Dnsmasq & DNS & Binary & v2.73rc6 & 9 & 27.6k\\
      TinyDTLS & DTLS & Binary & 06995d4 & 5 & 10.8k\\
      \bottomrule
      \end{tabular}
  \end{center}
\end{table}

\begin{table*}
  \caption{Average Accuracy of LLM Seed Interpretation}
  \label{table_accuracy}
  \begin{center}
  \begin{tabular}{lrrrrrrr}
      \toprule
      \textbf{Seed} & \textbf{Accuracy} & \textbf{Precision} & \textbf{\#G\_L} & \textbf{\#L\_MG} & \textbf{\#ML\_G} & \textbf{\#mismatch} & \textbf{\#field} \\
      \midrule
      dns\_queries            & 92.59\% & 66.67\% & 54.0   & 0.0   & 21.0   & 6.0     & 81.0   \\
      ssh            & 24.08\% & 18.03\% & 25.6 & 1.4 & 7.2  & 107.8 & 142.0  \\
      tls            & 70.96\% & 52.21\% & 28.4 & 0.2 & 10.0   & 15.8  & 54.4 \\
      dicom\_echo    & 45.22\% & 19.66\% & 14.0   & 1.0   & 17.2 &     39.0    & 71.2 \\
      dicom\_find    & 38.01\% & 15.37\% & 18.2 & 0.8 & 26.0   & 73.4  & 118.4 \\
      dicom\_get     & 19.61\% & 6.18\%  & 9.4  & 0.4 & 20.0   & 122.2 & 152.0 \\
      dicom\_store   & 22.93\% & 6.15\%  & 12.6 & 0.4 & 34.0   & 158.0   & 205.0 \\
          % ecc\_handshake\_client  & - & - & - & - & - & - & 107.4 \\
          % ecc\_handshake\_server  & - & - & - & - & - & - & 96.6 \\
          % psk\_handshake\_client  & - & - & - & - & - & - & 63.6 \\
          % psk\_handshake\_server  & - & - & - & - & - & - & 51.8 \\
      \bottomrule
  \end{tabular}   
  \end{center}
\end{table*}

%-------------------------------------------------------------------------------
\section{Evaluation}
\label{sec:evaluation}
%-------------------------------------------------------------------------------

In this study, we assess the capabilities of APFuzz by performing extensive experiments based on an evaluation benchmark for protocol fuzzing. The experimental plan aims to investigate the following research questions:

\begin{itemize}
    %\item[$\bullet$] \textbf{RQ1: How APFuzz compares to state-of-the-art protocol fuzzers?} We evaluate them with regard to both effectiveness and efficiency. Effectiveness is measured by code-coverage, state-coverage, and the crashes triggered on the targets. Efficiency is measured by the fuzzing throughput in terms of both executions and messages sent.
    %\item[$\bullet$] \textbf{RQ2: How APFuzz performs on handling binary protocol messages?} We look closer to the binary protocols to evaluate the effectiveness of LLM guidance in APFuzz, in terms of field-based binary message mutation as well as binary message generation for seed enriching.
    %\item[$\bullet$] \textbf{RQ3: What is the impact of different state model inference approaches?} We examine the state models obtained in different approaches from two aspects. One is to compare them to the ground-truth state models wherever available. The other is to compare their effectiveness on guiding the fuzzing process.
    \item[$\bullet$] \textbf{RQ1: Can we correctly interpret the binary message format of the seeds by using LLM?} We first look into the LLM-based input structure learning for binary protocols and evaluate the accuracy by comparing the fields derived by LLM with manually derived ground truth.
    \item[$\bullet$] \textbf{RQ2: Can we identify state variables automatically?} We compare the state variables identified in existing fuzzers and APFuzz and demonstrate the statistics about assigned values for the state variables during dynamic analysis.
    \item[$\bullet$] \textbf{RQ3: How does APFuzz compare to the baseline and state-of-the-art protocol fuzzers?} We evaluate them with regard to both effectiveness and efficiency. Effectiveness is measured by code coverage and the unique crashes triggered on the targets. Efficiency is measured by the fuzzing throughput in terms of both executions and messages sent per second.
\end{itemize}

\subsection{Experimental Setup}

To ensure reproducible results, we follow the fuzzing evaluation guidelines suggested in \cite{Schloegel2024SoK} to conduct experiments and document the process.

To allow comparisons of APFuzz and state-of-the-art protocol fuzzers under standardized conditions, we use ProFuzzBench \cite{Natella2021ProFuzzBench,Bohme2022benchmarking}, a widely adopted benchmark for protocol fuzzing. It supports a representative set of 13 network servers from popular open-source projects, covering C/C++ implementations of 10 application-layer protocols running over TCP and/or UDP. Specifically, half of the included protocols are text-based, while the other half are binary. Besides, from the perspective of implementation, these servers are diverse as they use a variety of APIs and fundamentally different mechanisms to handle client sessions and requests. For example, different methods are used for maintaining concurrency, e.g., pthread\_create() and fork(), and different function calls are adopted for network I/O, e.g., send(), receive(), fwrite(), fread(), evttp\_send\_reply(), and evhttp\_request\_get\_uri. A summary of the target servers and protocols in the benchmark is presented in Table \ref{table_benchmark}. 

\begin{table}[h]
  \caption{Static Analysis Results in Different Subject}
  \label{table_static_analysis}
  \centering
  \begin{tabular}{@{}clcccc@{}}
  \toprule
  \textbf{Target} & \textbf{Type} & \textbf{NS} & \textbf{SG} & \multicolumn{1}{l}{\textbf{SG*}} & \multicolumn{1}{l}{\textbf{Manual}} \\ \midrule
  Bftpd           & text          & 6           & 0           & 7                                & 4                                   \\
  Exim            & text          & 57          & 15          & 419                              & 8                                   \\
  Forked-daapd    & text          & 0           & 35          & 299                              & 4                                   \\
  LightFTP        & text          & 1           & 0           & 0                                & 1                                   \\
  Live555         & text          & 0           & 9           & 382                              & 3                                   \\
  ProFTPD         & text          & 81          & 18          & 819                              & 5                                   \\
  Pure-FTPd       & text          & 22          & 4           & 108                              & 3                                   \\
  Kamailio        & text          & 58          & 216         & 2723                             & 5                                   \\
  Dcmtk           & binary        & 0           & 388         & 2027                             & 5                                   \\
  OpenSSH         & binary        & 69          & 19          & 458                              & 7                                   \\
  OpenSSL         & binary        & 0           & 66          & 2263                             & 3                                   \\
  Dnsmasq         & binary        & 15          & 0           & 0                                & 5                                   \\
  TinyDTLS        & binary        & 4           & 5           & 14                               & 2                                   \\ \hline
  \end{tabular}
\end{table}

\begin{table}[]
  \caption{The Number of Identified State Variables}
  \centering
  \label{table_number_state_variables}
  \resizebox{\linewidth}{!}{
  \begin{tabular}{cccccl}
  \toprule
  \textbf{Target} & \multicolumn{1}{l}{\textbf{Type}} & \textbf{After Filter} & \multicolumn{1}{c}{\textbf{Example}} & \multicolumn{1}{l}{\textbf{Manual\cite{Qin2023NSFuzz}}} & \textbf{Example}         \\ \midrule
  Bftpd           & text                              & 1                     & state                                 & 4                                                                         & state           \\
  Exim            & text                              & 2                     & chunking\_state                       & 8                                                                         & chunking\_state \\
  Forked-daapd    & text                              & 0                     & -                                     & 4                                                                         & state           \\
  Live555         & text                              & 0                     & -                                     & 3                                                                         & fIsActive       \\
  ProFTPD         & text                              & 16                    & mode                                  & 5                                                                         & logged\_in      \\
  Pure-FTPd       & text                              & 3                     & state\_needs\_update                  & 3                                                                         & loggedin        \\
  OpenSSH         & binary                            & 7                     & session\_state                        & 7                                                                         & istate          \\
  OpenSSL         & binary                            & 3                     & ourpeer                               & 3                                                                         & hand\_state     \\
  Dnsmasq         & binary                            & 0                     & -                                     & 5                                                                         & found           \\
  TinyDTLS        & binary                            & 1                     & state                                 & 2                                                                         & state           \\ \cline{1-6}
  \end{tabular}
  }
\end{table}

\begin{figure*}
  \centering
  \subfigure{
  \begin{minipage}[b]{0.3\textwidth}
  \includegraphics[width=1\textwidth]{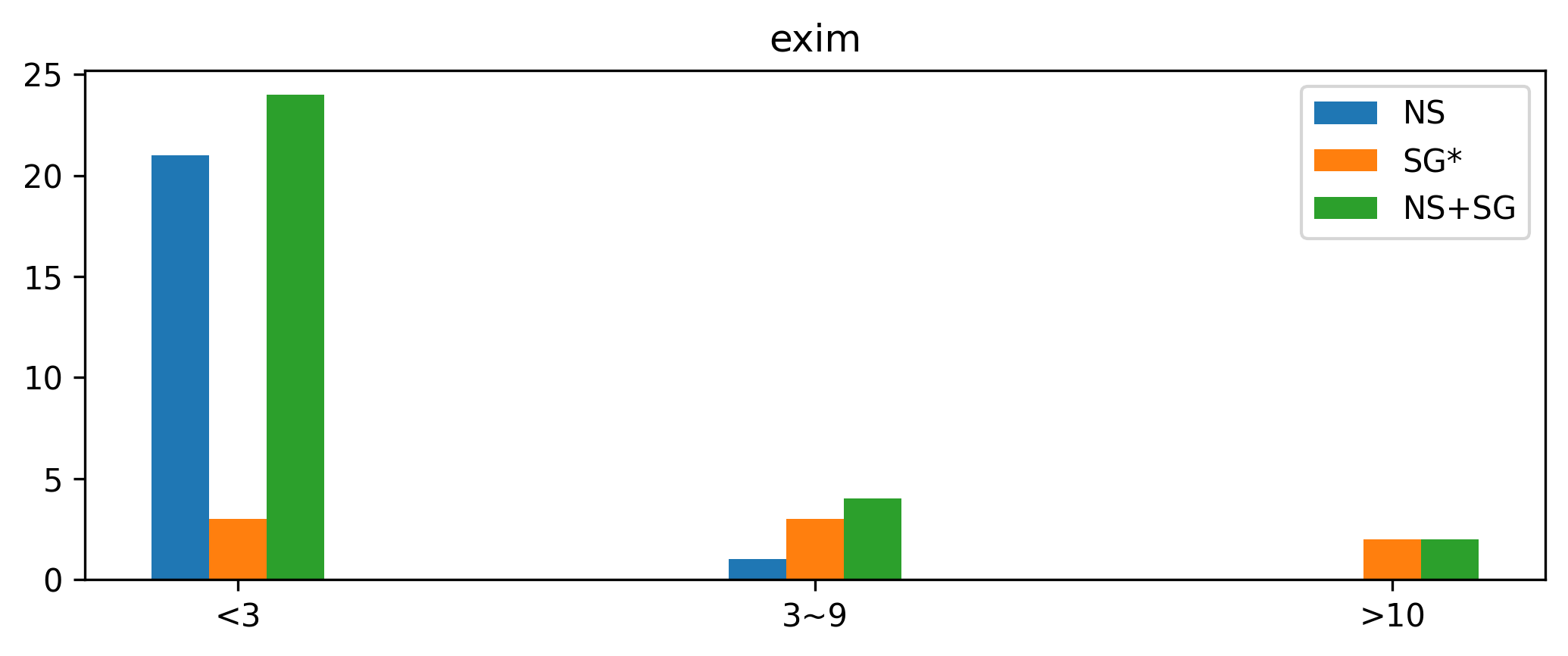} \\
  \includegraphics[width=1\textwidth]{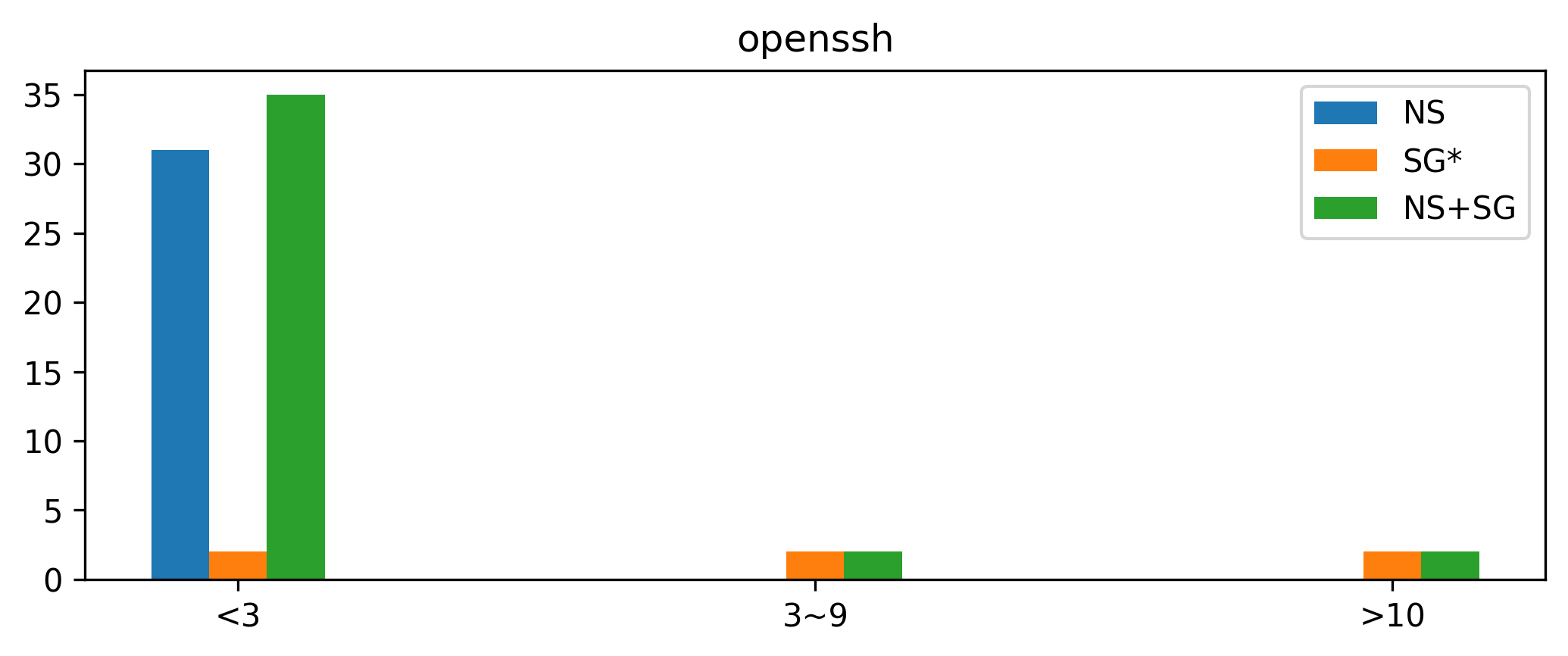} \\
  \includegraphics[width=1\textwidth]{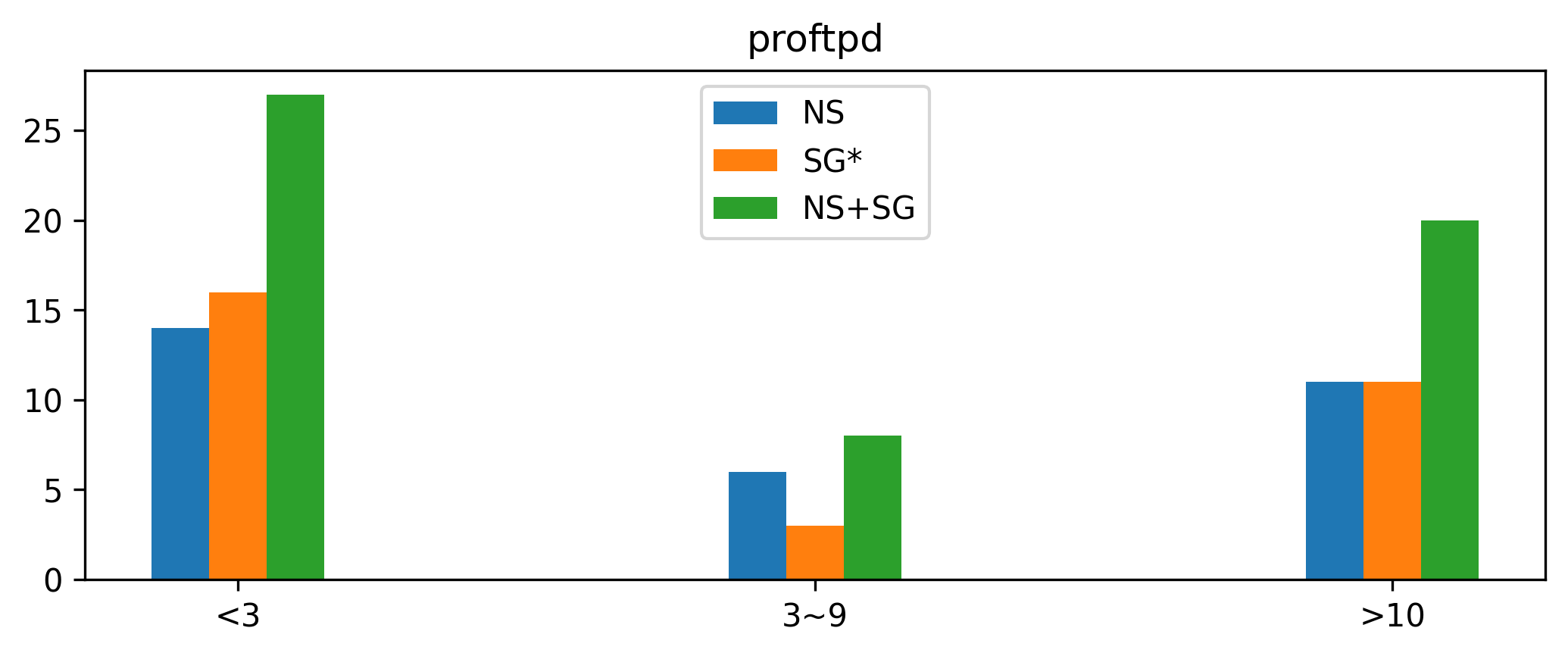}
  \end{minipage}
  }
  \subfigure{
  \begin{minipage}[b]{0.3\textwidth}
  \includegraphics[width=1\textwidth]{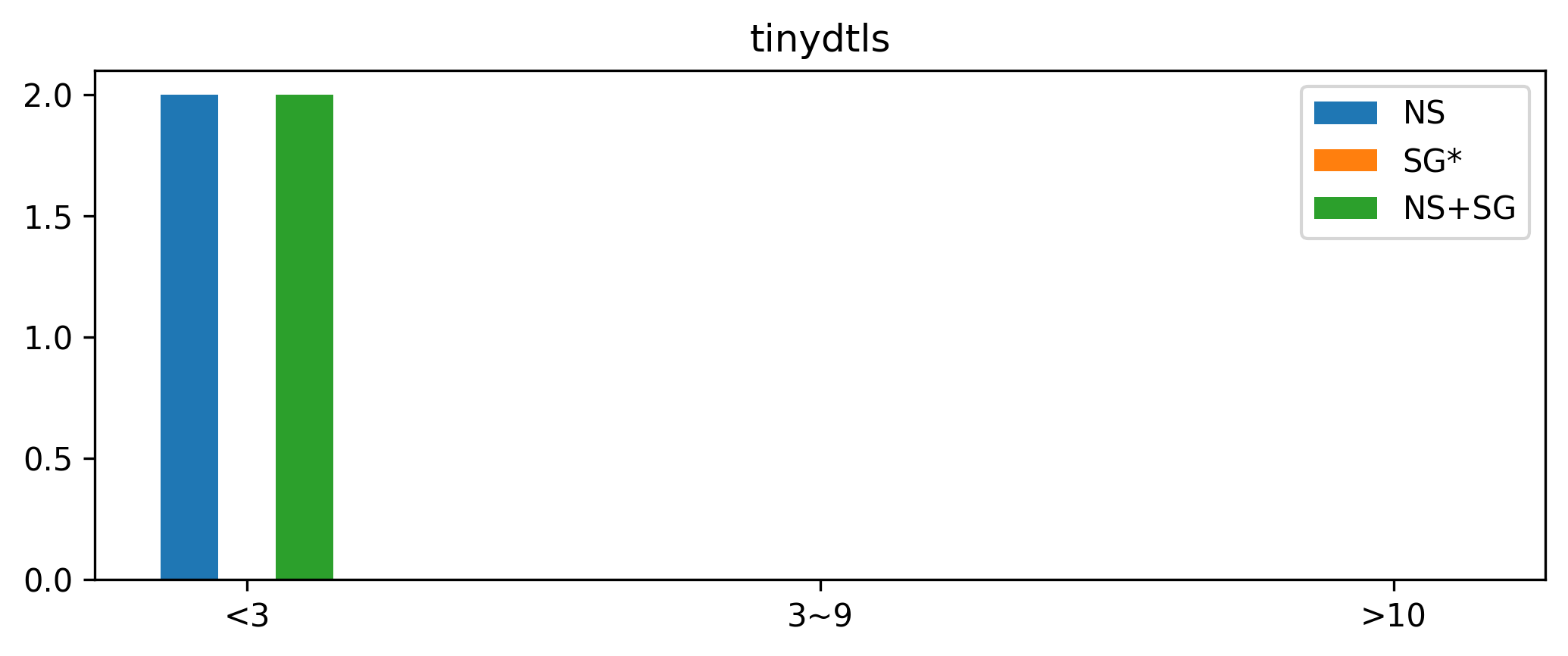} \\
  \includegraphics[width=1\textwidth]{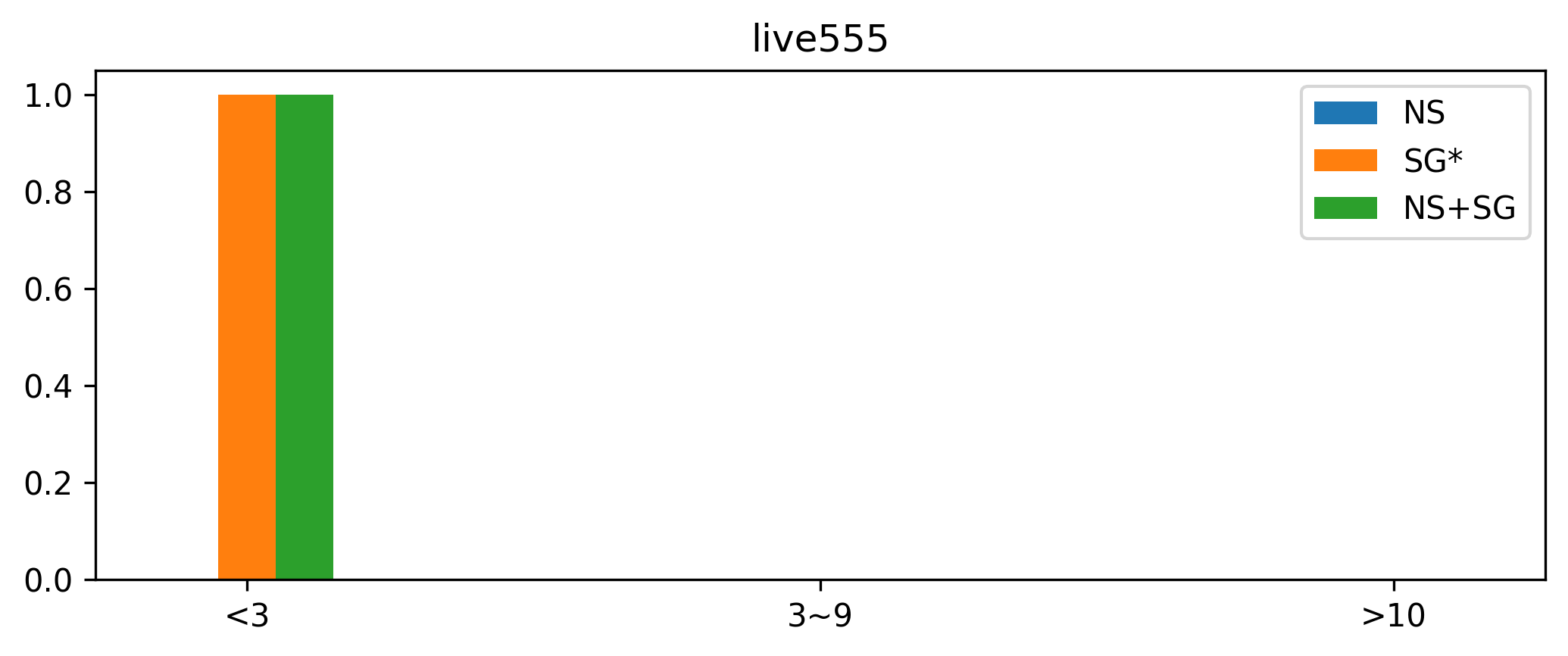} \\
  \includegraphics[width=1\textwidth]{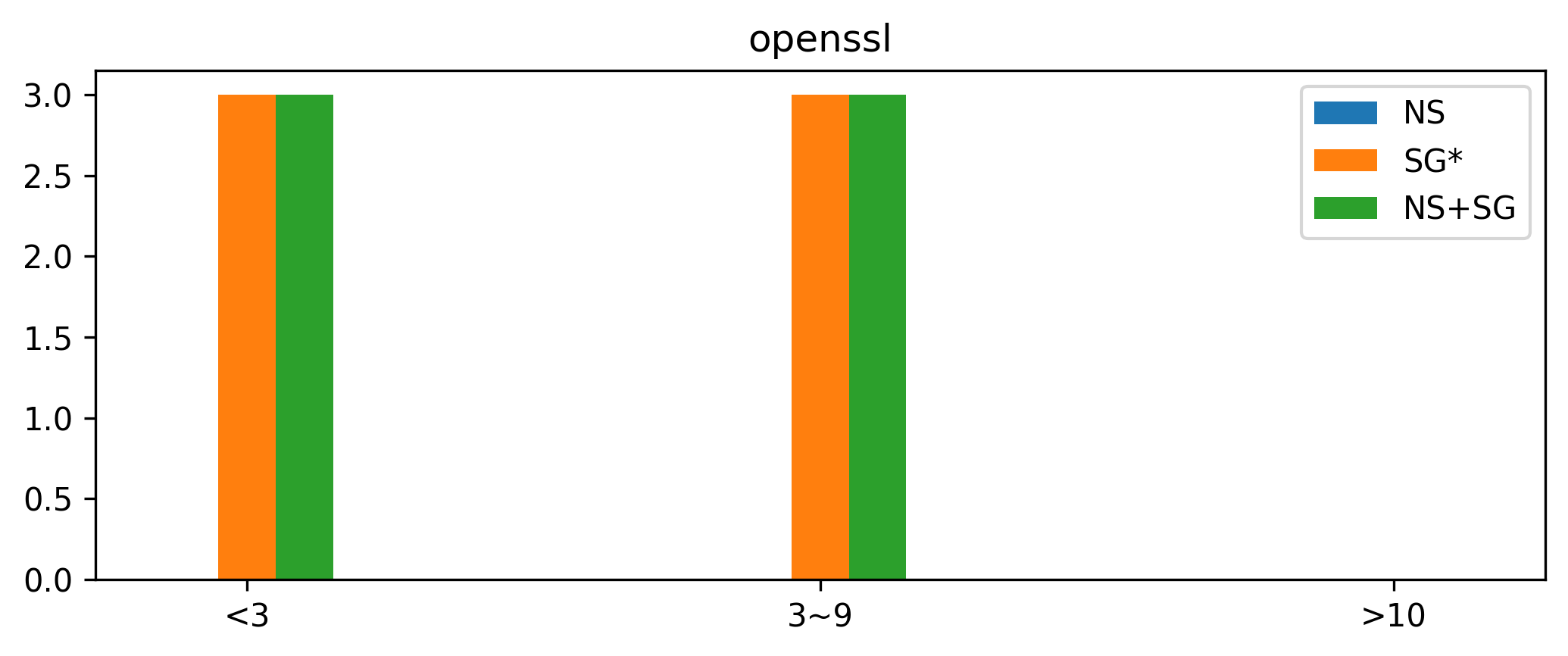}
  \end{minipage}
  }
  \subfigure{
  \begin{minipage}[b]{0.3\textwidth}
  \includegraphics[width=1\textwidth]{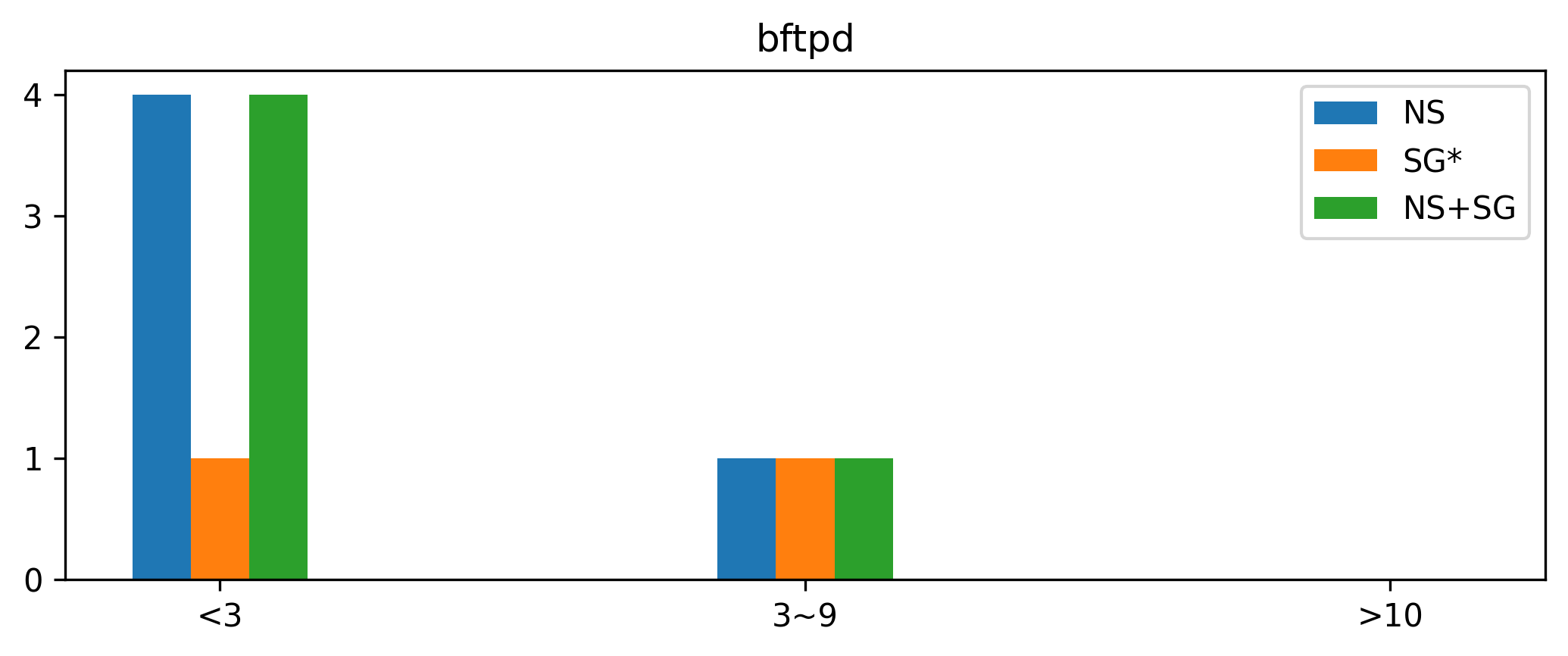} \\
  \includegraphics[width=1\textwidth]{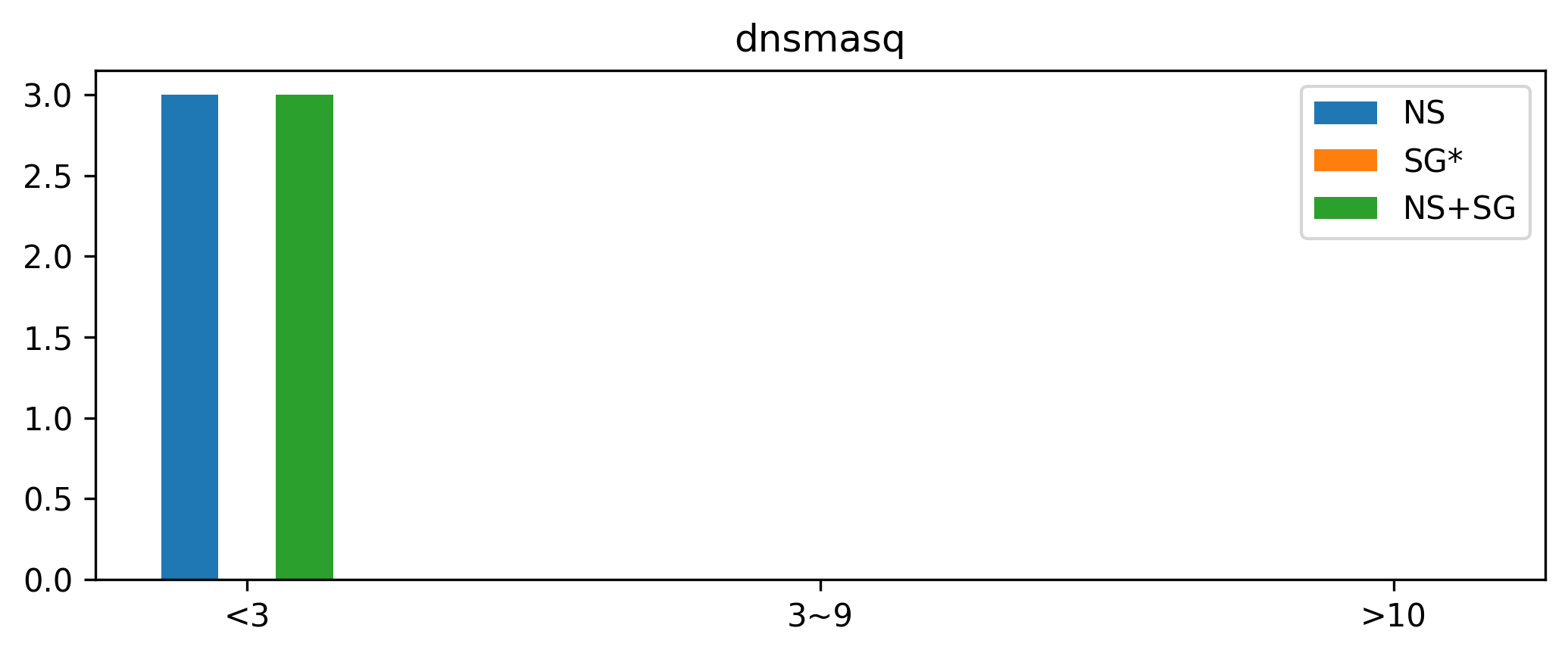} \\
  \includegraphics[width=1\textwidth]{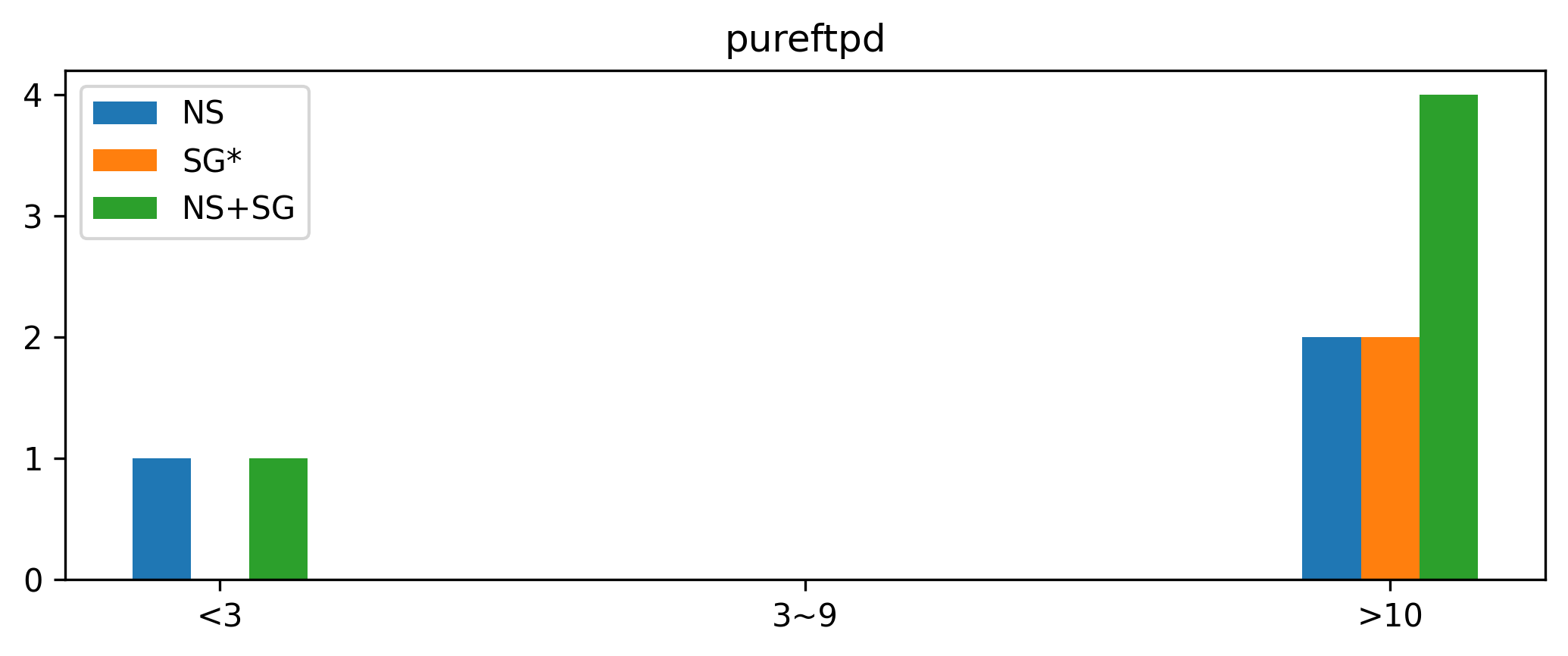}
  \end{minipage}
  }
  \caption{Changes in candidate state variables during 30 minutes of fuzzing. The vertical axis represents the number of variables, and the horizontal axis represents the number of unique values}
  \label{fig_dynamic_results}
\end{figure*}

ProFuzzBench provides a set of Dockerfiles for automatically building Docker images in which the target servers are set up and ready for fuzzing executions. The setup includes both best-practice server configurations for fuzzing and any necessary patches to the servers. For example, the patches remove randomization in the code to achieve consistent and reproducible server behavior (i.e., the same execution path upon the same input), which is the essential requirement of coverage-based fuzzing. ProFuzzBench also provides a set of initial seeds for the target servers, which are contributed by both researchers and practitioners involved in the benchmark project. The seeds are transformed from real-world network traffic trace that captures some typical usage patterns of the target servers, including some frequently used protocol commands and some valid credentials to get through the authentication process. We adopt these setups and initial seeds in our experiments.

We compare APFuzz to the baseline fuzzer AFLNET \cite{Pham2020AFLNET}, as well as the state-of-the-art greybox protocol fuzzers, including StateAFL \cite{Natella2022StateAFL}, NSFuzz \cite{Qin2023NSFuzz}, and ChatAFL \cite{Meng2024ChatAFL}. We consider AFLNET as the baseline fuzzer, as all other fuzzers in consideration are developed on top of it. Nonetheless, these fuzzers represent different directions of improvement over AFLNET. First, StateAFL addresses the issue of automatic state inferring and thus becomes a fully automated protocol fuzzing tool. Second, ChatAFL introduces the usage of LLMs to make the fuzzer smarter in terms of protocol awareness. Third, NSFuzz performs static analysis to identify state variables and client-handling loop entries, based on which it can achieve accurate state identification and fast synchronization. 

\begin{table*}
  \caption{Average Code Coverage Improvements of Different Fuzzers}
  \label{table_coverage}
  \begin{center}
  \resizebox{\textwidth}{!}{%
  \begin{tabular}{cc|rrrrrrr|rrrrrr}
      \toprule
      \multirow{2}{*}{\textbf{Target Service}} & \multirow{2}{*}{\textbf{Type}} & \multicolumn{7}{c|}{\textbf{Branch Coverage}} & \multicolumn{6}{c}{\textbf{Line Coverage}} \\
      \cmidrule(lr){3-15} &&   \textbf{AFLNET} &   \textbf{StateAFL}   &   \textbf{NSFuzz} &   \textbf{ChatAFL}    & \textbf{HNPFuzzer} & \textbf{SGFuzz} &  \textbf{APFuzz} &   \textbf{AFLNET} &   \textbf{StateAFL}   &   \textbf{NSFuzz} &   \textbf{ChatAFL}   &\textbf{HNPFuzzer} &   \textbf{APFuzz} \\             
      \midrule
      \textbf{Bftpd}&Text&483.6&-5.25\%&-1.57\%&-0.12\%&-1.36\%&--&-1.16\%&1253.6&-2.97\%&-1.40\%&+0.14\%&-0.34\%&-0.37\%\\
      \textbf{Exim}&Text&2904.2&+42.02\%&+44.23\%&+4.00\%&+4.66\%&--&+46.03\%&4772.6&+43.36\%&+45.52\%&+4.14\%&+4.26\%&+45.84\%     \\
      \textbf{Forked-daapd}&Text&2404.0&-5.61\%&+18.54\%&+11.64\%&+5.57\%&--&+15.34\%&7552.4&-5.90\%&+12.82\%&+6.72\%&+7.77  \%&+10.01\%\\
      \textbf{LightFTP}&Text&339.6&-19.55\%&+8.83\%&+8.48\%&+0.82\%&--&+7.60\%&704.2&-15.39\%&+4.63\%&+5.71\%&+1.51\%&+5.31\%  \\
      \textbf{Live555}&Text&2798.8&-0.56\%&-0.01\%&-0.54\%&-2.69\%&-38.86\%&+0.92\%&5630.4&-0.35\%&+0.40\%&-0.40\%&-1.83\%&+0.18\%     \\
      \textbf{ProFTPD}&Text&5161.6&-1.19\%&+0.90\%&-6.66\%&-0.34\%&--&+1.84\%&11176.2&-0.59\%&+0.74\%&-5.61\%&-0.27\%&+1.28\%\\
      \textbf{Pure-FTPd} &Text&1146.8&-8.02\%&+13.41\%&-9.64\%&+6.17\%&--&+7.94\%&2045.6&-5.75\%&+9.22\%&-6.55\%&+4.78\%&+5.82\%\\
      \textbf{Kamailio}&Text&9220.4&-7.92\%&+5.31\%&-24.21\%&-16.72\%&--&+2.93\%&15391.8&-3.36\%&+3.30\%&-11.17\%&-8.49\%&+2.25\%\\
      \textbf{Dcmtk}&Binary&3081.2&+0.45\%&-2.04\%&--&+0.67\%&--&-1.99\%&4162.6&+0.34\%&-1.41\%&--&+0.35\%&-1.19\%\\
      \textbf{OpenSSH}&Binary&3359.6&-1.27\%&-0.17\%&--&-2.34\%&--&-2.08&7529.2&-1.11\%&-0.29\%&--&-1.24\%&-1.64\%\\
      \textbf{OpenSSL}&Binary&10081.6&-0.37\%&+1.19\%&--&+1.06\%&-32.96\%&+1.20\%&23123.0&-0.12\%&+0.83\%&--&+0.70\%&+0.77\%\\
      \textbf{Dnsmasq}&Binary&906.4&+5.27\%&+5.98\%&--&+34.00\%&--&+35.86\%&1456.2&+4.24\%&+4.33\%&--&+27.88\%&+29.49\%\\
      \textbf{TinyDTLS}&Binary&561.6&-2.39\%&+16.74\%&--&+11.15\%&--&+15.92\%&1353.0&-4.02\%&+11.71\%&--&+5.99\%&+11.75\%\\
      \midrule
      \textbf{Average}&&-&-0.34\%&+8.56\%&-2.13\%&+3.13\%&-35.91\%&+10.03\%&-&+0.64\%&+6.98\%&-0.88\%&+3.16\%&+8.42\%\\
      \bottomrule
  \end{tabular}%
  }
  \end{center}
\end{table*}

% HNPFuzzer \cite{Fu2023HNPFuzzer},
% HNPFuzzer boosts the performance of AFLNET by eliminating the delays caused by network socket I/O, timeout-based synchronization and unnecessary server resets.

The hardware that we use to run the experiments has two Intel(R) Xeon(R) Platinum 8458P CPUs (each has 44 2.7GHz cores) and 3072 GB of memory with NVME SSD hard drives. Each fuzzing campaign is run in a standalone Docker container. All containers are allocated with precisely the same amount of computation resources, i.e., one CPU core with unlimited shared host memory and storage. We run 5 trials of 24-hour runtime for each fuzzer with each target server. 
%The number of resulted CPU hours of our fuzzing evaluation is TBA in total. 
It is also worth noting that most of the protocol fuzzers in consideration require some kind of pre-computations or manual preparations that take extra time. On the one hand, ChatAFL and APFuzz take additional time to communicate with ChatGPT before fuzzing for the purpose of pattern generation and seed enrichment. On the other hand, AFLNET and HNPFuzzer require manual instructions for protocol-specific status code interpretations, while NSFuzz requires manual annotations to accurately identify loop entries and state variables. The only exception is StateAFL, which uses a fully automatic approach that requires no preparations before fuzzing. Following the experimental approaches used in existing studies, the extra time is excluded from the 24-hour runtime of actual fuzzing. However, we record and report the additional time taken by ChatAFL and APFuzz in their pre-computations to facilitate fair comparisons.

\subsection{Input Structure Learning (RQ1)}

We use LLM to interpret the initial seeds of five binary protocol implementations in ProFuzzBench. For the purpose of evaluation, we build the ground truth by manually inspecting the messages in the seeds based on Wireshark as well as protocol specifications. The established ground truth includes the start bit, bit length, and field name for each field in a message. ProFuzzBench provides one seed for DNS, SSH, TLS, DTLS and four seeds for DICOM. We note that the pcap file of the seed for DTLS is missing, thus we are unable to establish its ground truth.
%In a typical fuzzing process, the mutation of seeds is usually completely random. Since binary protocol seeds involve a large number of fields, if certain bytes are randomly selected for mutation, most of these mutations would be ineffective. By using LLM to interpret the seeds, it is possible to accurately locate the bit positions of the fields, which provides strong guidance for fuzzing binary protocols. 

%Establish the ground truth for the initial seeds of binary protocols, except for the DTLS protocol, and use LLM to interpret the seed files, creating a mechanism to evaluate the accuracy of LLM's interpretation. 

To measure the accuracy of seed interpretation, we match the fields identified by LLM to the fields in the ground truth. There are three types of field matching. 
$\#L\_G$ refers to the number of cases where an LLM-identified field precisely matches a ground-truth field. 
$\#L\_MG$ refers to the number of cases where an LLM-identified field matches multiple consecutive ground-truth fields.
$\#ML\_G$ refers to the number of cases where multiple consecutive LLM-identified fields match a single ground-truth field.

Based on different types of field matching, we calculate the accuracy rate of precise single-filed matching $exact\_acc$ and the accuracy rate that includes both single-filed matching and multiple-field matching $multi\_acc$ as follows.
\begin{equation}
exact\_acc = \frac{\#G\_L}{\#G\_L + \#MG\_L + \#G\_ML + \#mismatch}
\end{equation}
\begin{equation}
multi\_acc = \frac{\#G + \#MG\_L + \#G\_ML\_L}{\#G\_L + \#MG\_L + \#G\_ML + \#mismatch}
\end{equation}

%To more precisely measure the accuracy of LLM parsing, we consider various matching scenarios, which can be broadly categorized into single-field matching, Groundtruth multi-field combined matching with an LLM-parsed field, and LLM-parsed fields combined matching with a Groundtruth field. In the code logic, for each LLM field, we first attempt to match it with a single field in the Groundtruth. If the match is successful, $G\_L\_match$ is incremented. If single-field matching fails, the function tries to combine multiple Groundtruth fields and match them with a single LLM field. If successful, $MG\_L\_match$ is incremented. If both of the above matchings fail, the function then attempts to combine multiple LLM fields and match them with a single Groundtruth field. If successful, $G\_ML\_match$ is incremented, and the already matched LLM fields are skipped to avoid double-counting. If all matching methods fail, the field is counted as unmatched, and $mismatch$ is incremented. 

To minimize randomness, we use the average of five LLM parsing attempts to evaluate the accuracy, and the final results are shown in Table \ref{table_accuracy}. The parsing accuracy of the DNS protocol seeds is the highest, with the single-field matching accuracy reaching 66.67\% and the overall accuracy reaching an impressive 92.59\%. It can also be observed that due to the complexity of the SSH and DICOM protocol specifications, the accuracy of seed file parsing for these two protocols is generally low. However, the parsed results can provide valuable guidance for field-level mutations during the fuzzing process rather than entirely random mutations. We note that the parsing accuracy for some protocols may be limited, but incorrect information is quickly discarded during the early stages of the fuzzing process. In the worst case, the process reverts to traditional randomized fuzzing without significantly affecting the fuzzer's performance. However, even partial accuracy in parsing results proves valuable. Correctly parsed portions of the message format provide guidance, enabling the fuzzer to generate test cases that adhere to network protocol communication rules and improve the overall fuzzing efficiency of APFuzz.

\subsection{State Representation Learning (RQ2)}

In this part, we discuss the identification of state variables and experimentally demonstrate the feasibility of the identification method we adopted. This method consists of two parts: static analysis and dynamic analysis, both of which will be explored in detail.

\begin{table*}
  \caption{Average Throughput on Different Fuzzers with 5 Runs of 24 Hours}
  \label{table_throughput}
  \footnotesize
  \begin{center}
  \resizebox{\textwidth}{!}{
     \begin{tabular}{cc|rrrrrr|rrrrrr}
        \toprule
      \multirow{2}{*}{\textbf{Target Service}}& \multirow{2}{*}{\textbf{Type}}&\multicolumn{6}{c|}{\textbf{Executions per Second}} & \multicolumn{6}{c}{\textbf{Messages Sent per Second}} \\
      \cmidrule(lr){3-14}&
      & \textbf{AFLNET} & \textbf{StateAFL}  & \textbf{NSFuzz} & \textbf{ChatAFL} &\textbf{ HNPFuzzer} & \textbf{APFuzz} & \textbf{AFLNET} & \textbf{StateAFL}  & \textbf{NSFuzz} & \textbf{ChatAFL} & \textbf{HNPFuzzer} & \textbf{APFuzz} \\ 
      \midrule
      \textbf{Bftpd} &Text&3.49 &0.77$\times$ &10.40$\times$ & 1.18$\times$ &2.91$\times$ &10.49$\times$ &74.3 & 1.37$\times$ &6.51$\times$ &1.35$\times$ &2.46$\times$  &6.92$\times$ \\
       \textbf{Exim} &Text&2.72 &0.83$\times$ &2.78$\times$ &0.88$\times$ &2.76$\times$ &2.96$\times$ &22.69 &0.66$\times$  &1.59$\times$  &1.39$\times$ &4.15$\times$ &3.08$\times$ \\
       \textbf{Forked-daapd} &Text&0.73 &0.99$\times$ &4.10$\times$ &1.05$\times$	&3.47$\times$ &2.84$\times$ &6.13 &0.75$\times$ &1.35$\times$&1.92$\times$ &1.63$\times$ &1.54$\times$ \\
       \textbf{LightFTP}&Text &2.56 &0.72$\times$ &12.33$\times$ &1.30$\times$ &7.68$\times$ &7.22$\times$ &75.87 &0.61$\times$ &4.42$\times$ &1.42$\times$ &11.66$\times$ &3.49$\times$\\
       \textbf{Live555} &Text&8.14 &0.55$\times$&6.60$\times$&1.04$\times$ &0.58$\times$ &3.77$\times$	&116.59&0.63$\times$&8.27$\times$ &1.18$\times$ &0.53$\times$&5.94$\times$\\
       \textbf{ProFTPD}&Text &2.47 &0.74$\times$	&4.06$\times$ &0.70$\times$&1.68$\times$ &6.51$\times$&77.94 & 0.88$\times$&6.13$\times$ &0.83$\times$&0.97$\times$ &8.04$\times$ \\
       \textbf{Pure-FTPd} &Text&4.58 &	0.87$\times$&	5.00$\times$&	1.71$\times$&	2.66$\times$&	4.22$\times$&	99.69&	0.67$\times$&	3.11$\times$&	2.45$\times$&	2.33$\times$&	9.93$\times$ \\
       \textbf{Kamailio}&Text &3.59&	1.19$\times$&	1.87$\times$&	1.02$\times$&	2.35$\times$&	1.86$\times$&32.5	&1.12$\times$&	1.70$\times$&	2.01$\times$&	4.28$\times$&	1.59$\times$\\
      \textbf{Dcmtk}&Binary &13.38&	1.51$\times$&	1.14$\times$&	--&	1.12$\times$&	0.95$\times$&	43.62&	1.27$\times$&	2.00$\times$&	--&	0.81$\times$&	2.00$\times$\\
      \textbf{OpenSSH}&Binary  & 14.46&	0.50$\times$&	1.48$\times$&	--&	1.11$\times$&	1.39$\times$&	50.96&	0.60$\times$&	6.16$\times$&	--&	0.64$\times$&	4.96$\times$ \\
      \textbf{OpenSSL}&Binary  & 5.33 &	0.83$\times$&	2.49$\times$&	--&	2.58$\times$&	2.12$\times$&	38.97	&1.03$\times$&	0.85$\times$&	--&	5.57$\times$&	1.57$\times$ \\
      \textbf{Dnsmasq}&Binary  & 4.62 &1.20$\times$&	9.60$\times$&	--&	3.87$\times$&	6.29$\times$&	144.22&	0.92$\times$&	10.75$\times$&	--&	5.06$\times$&	8.10$\times$\\
      \textbf{TinyDTLS} &Binary & 2.4	&0.7$\times$&	31.80$\times$&	--&	30.68$\times$&	16.86$\times$&	35.81&	1.01$\times$&	42.72$\times$&	--&	82.12$\times$&	23.48$\times$\\
      \midrule
        \textbf{Average}&&-&0.88$\times$&7.20$\times$&1.11$\times$&4.88$\times$&5.19$\times$&-&0.89$\times$&7.29$\times$&1.57$\times$&9.40$\times$&6.20$\times$\\
      \bottomrule
      \end{tabular}
  }
  \end{center}
\end{table*}

\begin{table*}
  \caption{Average Number of Vertexes and Edges of the Fuzzer-inferred State Models}
  \label{table_state}
  \footnotesize
  \begin{center}
  \resizebox{\textwidth} {!}{
  \begin{tabular}{ccrrrrrrrrrrrr}
      \toprule
      \multirow{2}{*}{\textbf{Target Service}} &\multirow{2}{*}{\textbf{Type}}& \multicolumn{2}{c}{\textbf{AFLNET}} & \multicolumn{2}{c}{\textbf{StateAFL}} & \multicolumn{2}{c}{\textbf{NSFuzz}} & \multicolumn{2}{c}{\textbf{ChatAFL}}& \multicolumn{2}{c}{\textbf{HNPFuzzer}} & \multicolumn{2}{c}{\textbf{APFuzz}}\\ 
      \cmidrule(lr){3-14}&& \textbf{Vertexes} & \textbf{Edges} & \textbf{Vertexes} & \textbf{Edges} &\textbf{Vertexes} & \textbf{Edges} &\textbf{Vertexes} & \textbf{Edges} &\textbf{Vertexes} & \textbf{Edges}&\textbf{Vertexes} & \textbf{Edges}\\
      \midrule
       \textbf{Bftpd}&Text &24.0&175.4&4.6&8.2&14.2&41.6&24.0&259.4&23.0&224.6&5.0&11.0 \\
       \textbf{Exim} &Text&11.0&50.6&44.6&104&5.6&15.0&13.6&83.8&11.2&68.0&4.2&8.6 \\
       \textbf{Forked-daapd}&Text &8.2&22.2&3.4&3.8&1.0&1.0&8.0&23.8&8.0&28.4&1.0&1.0\\
       \textbf{LightFTP}&Text&23.8&184.0&8.4&17.4&4.0&8.0&24.2&259.8&22.6&252.8&4.0&8.0\\
       \textbf{Live555}&Text &10.0&72.6&17.2&42.2&15.0&52.2&11.6&109.2&10.0&80.8&14.6&53.2 \\
       \textbf{ProFTPD} &Text&28.0&255.0&5.4&12.8&29.6&137.2&27.4&325.0&29.6&319.8&32.4&160.4\\
       \textbf{Pure-FTPd}&Text &28.8&273.0&6.2&13.4&6.0&18.2&28.8&310.2&28.6&383.8&74.4&944.0\\
       \textbf{Kamailio}&Text &13.4&96.0&5.6&10.4&4.6&12.6&14.2&125.2&10.2&83.6&4.8&13.2 \\
      \textbf{Dcmtk}&Binary & 4.0 & 3.0 & 21.8 & 41.6 & 6.8 & 13.6  & -- & -- & 6.0& 6.0 & 7.2 & 15.0 \\
      \textbf{OpenSSH}&Binary  & 48.6 & 90.4 & 82.2 & 228.6 & 20.0 & 47.0 & -- & -- & 24.0 & 71.2& 14.0 & 22.6  \\
      \textbf{OpenSSL}&Binary  & 68.8 & 73.6 & 2.2 & 2.4 & 17.2 & 30.4 & -- & -- & 8.0 & 15.4 & 9.2 & 13.4  \\
      \textbf{Dnsmasq}&Binary  & 106.0 & 334.0 & 154.0 & 756.8 & 2.0 & 2.0 & -- & -- & 283.4 & 1266.4 & 1.0 & 1.0  \\
      \textbf{TinyDTLS} &Binary & 8.8 & 25.6 & 7.8 & 16.4 & 7.0 & 26.6 & -- & -- & 9.4 & 52.6 & 1.0 & 1.0 \\    
      \bottomrule
  \end{tabular}
  }
  \end{center}
\end{table*}

We compare three static analysis strategies: NSFuzz (NS), SGFuzz (SG), and extended SGFuzz* (SG*) against manual annotation. Table \ref{table_static_analysis} shows SG* identifies significantly more state variables than NS due to broader enum-type matching rules, but with insufficient accuracy. NS more precisely identifies communication-related state variables but still detects excessive redundancy compared to manual labeling.

To address this, we conducted dynamic analysis using AFLNET across ProFuzzBench targets for 30 minutes, analyzing unique values for each candidate variable. Figure \ref{fig_dynamic_results} reveals that despite large numbers of statically identified variables, only small subsets actually change during fuzzing. NS-identified variables are generally more fuzzing-relevant, though SG* detected variables in Live555 and OpenSSL where NS failed. APFuzz combines both techniques for comprehensive identification.

Based on dynamic analysis, we established filtering rules: select variables with 3-9 unique values, exclude those exceeding 10,000 unique values, and prioritize variables containing "state" in their names. Variables with fewer than 3 values show minimal fuzzing impact, while those exceeding 3 values can create oversensitive state models.

Table \ref{table_number_state_variables} shows our combined method identifies state variables similar to manual annotation with high confidence across various ProFuzzBench targets. For targets where automatic identification failed (likely due to insufficient dynamic analysis time), we used NSFuzz's manual annotation approach.

\begin{figure}
  \begin{center}
      \includegraphics[width=0.5\linewidth]{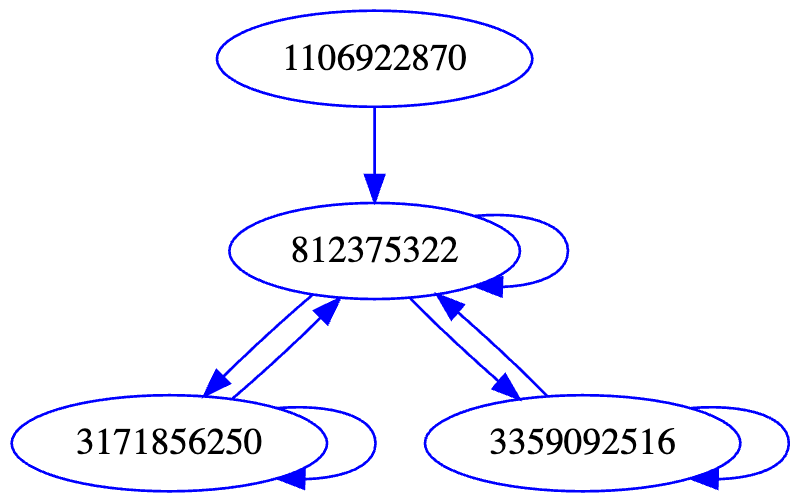}
  \end{center}
  \caption{The inferred state model of LightFTP by APFuzz}
  \label{fig:lightftp-APFuzz}
  \end{figure}
  
  \begin{figure}
  \begin{center}
      \includegraphics[width=1\linewidth]{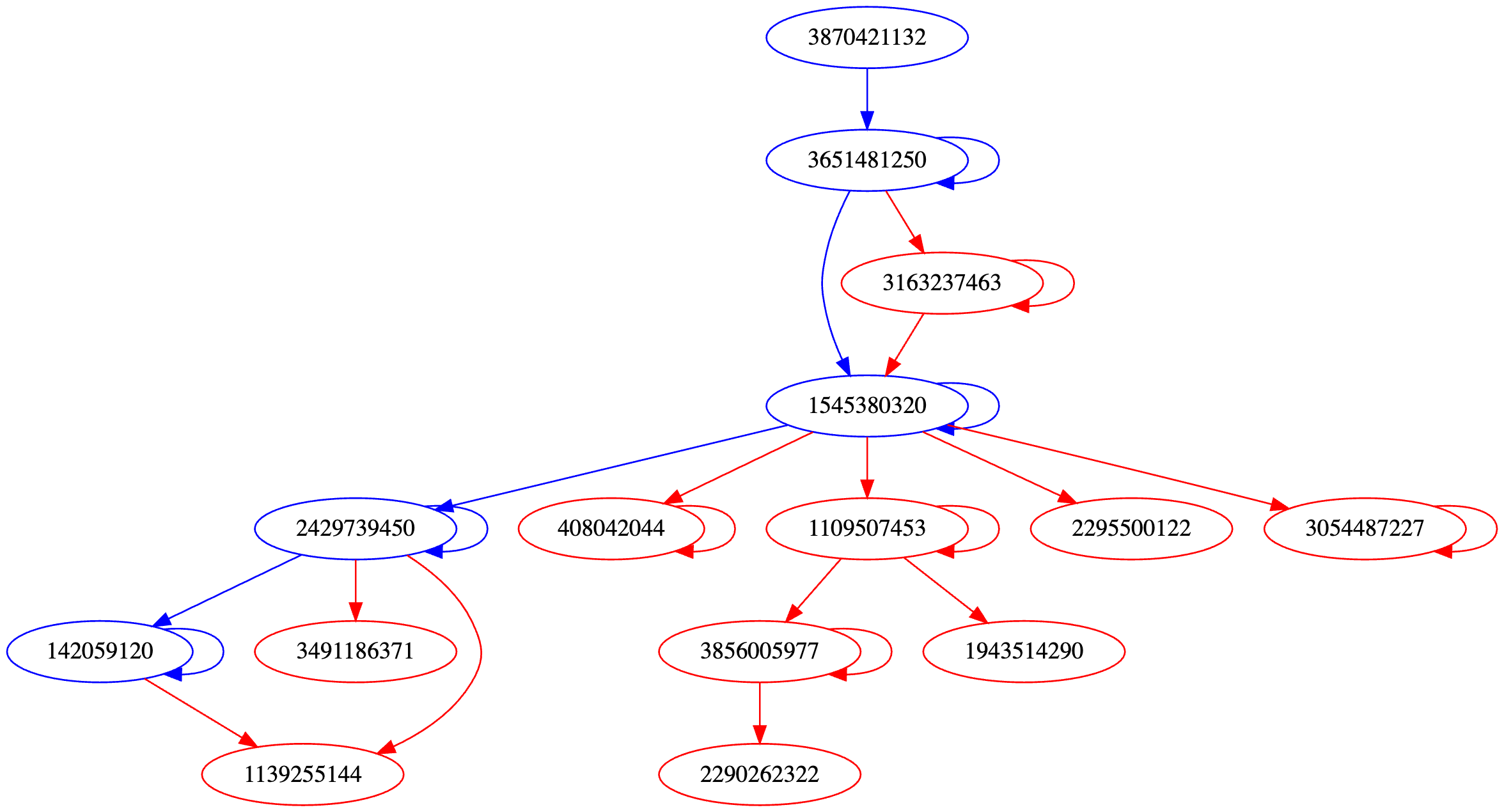}
  \end{center}
  \caption{The inferred state model of OpenSSH by APFuzz}
  \label{fig:openssh-APFuzz}
\end{figure}

\begin{table}[h]
  \caption{Number of Unique crashes Found by Different Fuzzers}
  \label{table:crashes-count}
  \resizebox{\linewidth}{!}{%
  \begin{tabular}{c|cccccc}
  \toprule
  \textbf{Subject} & \textbf{AFLNET} & \textbf{NSFuzz} & \textbf{APFuzz} & \textbf{StateAFL} & \textbf{HNPFuzzer} & \textbf{ChatAFL} \\ 
  \midrule
  Dnsmasq  & 1      & 3      & 3      & 3        & 3         & -       \\
  Tinydtls & 3      & 4      & 5      & 4        & 2         & -       \\
  Live555  & 5      & 5      & 5      & 5        & 3         & 5       \\ 
  \midrule
  Total    & 8      & 12     & 13     & 12       & 8         & 5       \\ 
  \bottomrule
  \end{tabular}%
  }
\end{table}

\begin{table*}[]
\caption{Comparison of Bug Detection Time Across Different Fuzzers (in seconds)}
\label{table:vul-time}
\begin{tabular}{ccc|cccccc}
\hline
\textbf{Target} & \textbf{Type}          & \textbf{Position}     & \textbf{AFLNET}   & \textbf{NSFuzz}   & \textbf{StateAFL} & \textbf{HNPFuzzer} & \textbf{ChatAFL}  & \textbf{APFuzz}   \\ \hline
Dnsmasq          & heap-buffer-overflow   & rfc1035.c:109         & \textgreater{}24h & \textgreater{}24h & 39409             & \textgreater{}24h  & -                 & \textbf{4246}     \\
Dnsmasq          & heap-buffer-overflow   & rfc1035.c:120         & \textgreater{}24h & \textgreater{}24h & \textgreater{}24h & \textgreater{}24h  & -                 & \textbf{5484}     \\
Dnsmasq          & heap-buffer-overflow   & rfc1035.c:152         & 22518             & 2477              & 8106              & 4595               & -                 & 4200              \\
Dnsmasq          & heap-buffer-overflow   & rfc1035.c:181         & \textgreater{}24h & \textgreater{}24h & 4863              & 8393               & -                 & \textgreater{}24h \\
Dnsmasq          & heap-buffer-overflow   & rfc1035.c:122         & \textgreater{}24h & \textgreater{}24h & \textgreater{}24h & 5856               & -                 & \textgreater{}24h \\
Tinydtls         & global-buffer-overflow & dtls-numeric.h:93     & 15589             & 17490             & 41251             & \textgreater{}24h  & -                 & \textbf{148}      \\
Tinydtls         & global-buffer-overflow & dtls-numeric.h:94     & 737               & \textgreater{}24h & 1445              & 971                & -                 & \textbf{702}      \\
Tinydtls         & global-buffer-overflow & dtls.c:361            & \textgreater{}24h & 1303              & \textgreater{}24h & \textgreater{}24h  & -                 & \textbf{559}      \\
Tinydtls         & global-buffer-overflow & sha2.c:494            & 24                & 57                & 77                & 5                  & -                 & 180               \\
Live555          & heap-use-after-free    & MPEG1or2Demux.cpp:129 & 4770              & 9441              & 2867              & 10884              & 2157              & \textbf{1196}     \\
Live555          & heap-use-after-free    & Media.h:59            & 1079              & 1705              & 1630              & 4299               & 1092              & \textbf{978}      \\
Live555          & stack-buffer-overflow  & RTSPServer.cpp:397    & \textgreater{}24h & \textgreater{}24h & \textgreater{}24h & \textgreater{}24h  & 65606             & \textgreater{}24h \\
Live555          & stack-use-after-return & RTSPServer.cpp:607    & \textgreater{}24h & 14195             & \textgreater{}24h & \textgreater{}24h  & \textgreater{}24h & 27883             \\ \hline
\end{tabular}
\end{table*}

\subsection{Fuzzing Performance (RQ3)}

%\tiny
%\scriptsize
%\footnotesize
%\small

We evaluate APFuzz's feasibility using code coverage as the standard metric for fuzzer effectiveness. Table \ref{table_coverage} shows average branch and line coverage across five experimental runs for 13 target services. APFuzz achieved the best results compared to state-of-the-art fuzzers, with an average increase of 10.03% in branch coverage and 8.42% in line coverage over AFLNET.
Table \ref{table_throughput} presents throughput statistics including message processing speed and execution speed. NSFuzz achieved the fastest message processing speed while HNPFuzzer had the fastest execution speed. APFuzz ranks among the faster fuzzers overall, providing sufficient throughput for comprehensive state exploration.
State model inference varies across fuzzers due to different representation methods: AFLNET uses server response codes, StateAFL utilizes fuzzy hashes of memory data, and NSFuzz adopts variable-based representations. Table \ref{table_state} shows the average vertices and edges in state models inferred during 24-hour experiments. Despite targeting the same services, different fuzzers generate distinct state models. Figures \ref{fig:lightftp-APFuzz} and \ref{fig:openssh-APFuzz} illustrate APFuzz's state models for LightFTP and OpenSSH respectively.

\subsection{Potential Bug Finding Effectiveness (RQ3)}

To evaluate vulnerability discovery capabilities, we replayed crashing test cases from five independent 24-hour fuzzing runs and analyzed outputs. AddressSanitizer \cite{serebryany2012addresssanitizer} was used to deduplicate crashes by program addresses, ensuring fair comparison across fuzzers.

Table \ref{table:crashes-count} shows unique crashes identified by each fuzzer. APFuzz achieved the best performance with 13 unique crashes, followed by NSFuzz and StateAFL (12 each), while AFLNET and HNPFuzzer found 8 each. ChatAFL discovered 5 crashes in Live555. APFuzz consistently outperformed others across different targets, finding 5 unique crashes in TinyDTLS compared to 4 or fewer by other fuzzers.

Table \ref{table:vul-time} compares vulnerability detection times. APFuzz demonstrated superior efficiency, detecting heap-buffer-overflow vulnerabilities in Dnsmasq within 4,246 and 5,484 seconds, while other fuzzers took significantly longer or failed within the 24-hour period. In TinyDTLS, APFuzz quickly identified three global-buffer-overflow vulnerabilities (148, 702, and 559 seconds), and in Live555, it discovered heap-use-after-free vulnerabilities in 1,196 and 978 seconds.

APFuzz's ability to identify diverse memory safety bugs—including heap-buffer-overflow, global-buffer-overflow, heap-use-after-free, and stack-related issues—demonstrates its effectiveness in detecting critical vulnerabilities in network protocol implementations

\section{discussion}
\label{sec:discussion}

This paper contributes to the field of automatic protocol fuzzing systems through the proposal of APFuzz, a novel framework designed to enhance protocol fuzzing capabilities by leveraging large language models and a combination of static and dynamic analysis techniques. APFuzz captures message format information via large language models (LLMs), introducing an advanced method for understanding protocol structures and providing better insights into message formats without relying solely on manual analysis or heuristics. By combining static and dynamic methods, APFuzz identifies suitable state variables that reflect protocol status, enhancing effectiveness for fuzzing campaigns. However, the current implementation has limitations. It primarily focuses on message format extraction and does not fully address the dependency problem inherent in binary protocols, which impacts its ability to model and test protocols with complex dependencies. Additionally, the use of simple counting methods for filtering state variables may include irrelevant variables, as the filtering results are still partly influenced by the quality of static analysis, leading to potential noise. Despite these challenges, APFuzz set the stage toward automated and efficient protocol fuzzing. Future work could focus on addressing protocol field dependencies in binary protocols, improving state variable filtering with advanced network protocol reverse engineering methods, and expanding support for diverse protocols and testing scenarios.
%-------------------------------------------------------------------------------
%\section{Discussion}
%\label{sec:discussion}
%-------------------------------------------------------------------------------

%The discussion goes here.

%-------------------------------------------------------------------------------
\section{Conclusion}
%-------------------------------------------------------------------------------

In this paper, we proposed to enhance the smartness of greybox protocol fuzzers from the perspectives of both the state model and the message model of communication protocols. The proposed approaches are implemented in APFuzz. It supports automatic state representation learning and state model inference. It also adopts LLM-based binary message interpretations to enhance the structure awareness in message mutations. In the evaluations, we compare APFuzz with the baseline fuzzer AFLNET as well as the state-of-the-art fuzzers. The results show that APFuzz outperforms existing fuzzers, achieving 10.03\% and 8.42\% improvements over the baseline fuzzer on branch coverage and line coverage, respectively, with 5.19$\times$ and 6.20$\times$ execution and message-sending throughput in a fully automatic manner.

\section*{Acknowledgment}
This research was conducted under the 6G Security Research and Development Project led by the Commonwealth Scientific and Industrial Research Organisation (CSIRO), with funding appropriated by the Australian Government Department of Home Affairs. The views expressed in this paper are those of the authors and do not necessarily reflect any Australian Government policy position. More information about the Project is available at \url{https://research.csiro.au/6gsecurity/}.

\bibliographystyle{plain}
\bibliography{main}

@article{Qin2023NSFuzz,
author = {Qin, Shisong and Hu, Fan and Ma, Zheyu and Zhao, Bodong and Yin, Tingting and Zhang, Chao},
title = {NSFuzz: Towards Efficient and State-Aware Network Service Fuzzing},
year = {2023},
issue_date = {November 2023},
publisher = {Association for Computing Machinery},
address = {New York, NY, USA},
volume = {32},
number = {6},
issn = {1049-331X},
url = {https://doi.org/10.1145/3580598},
doi = {10.1145/3580598},
journal = {ACM Trans. Softw. Eng. Methodol.},
month = {sep},
articleno = {160},
numpages = {26}
}

@inproceedings {Ba2022SGFuzz,
author = {Jinsheng Ba and Marcel B{\"o}hme and Zahra Mirzamomen and Abhik Roychoudhury},
title = {Stateful Greybox Fuzzing},
booktitle = {31st USENIX Security Symposium (USENIX Security 22)},
year = {2022},
isbn = {978-1-939133-31-1},
address = {Boston, MA},
pages = {3255--3272},
url = {https://github.com/bajinsheng/SGFuzz.git},
publisher = {USENIX Association},
month = aug
}

@article{Natella2022StateAFL,
author = {Natella, Roberto},
title = {StateAFL: Greybox fuzzing for stateful network servers},
year = {2022},
issue_date = {Dec 2022},
publisher = {Kluwer Academic Publishers},
address = {USA},
volume = {27},
number = {7},
issn = {1382-3256},
url = {https://doi.org/10.1007/s10664-022-10233-3},
doi = {10.1007/s10664-022-10233-3},
journal = {Empirical Softw. Engg.},
month = {dec},
numpages = {31}
}

@ARTICLE{Li2022SNPSFuzzer,
  author={Li, Junqiang and Li, Senyi and Sun, Gang and Chen, Ting and Yu, Hongfang},
  journal={IEEE Transactions on Information Forensics and Security}, 
  title={SNPSFuzzer: A Fast Greybox Fuzzer for Stateful Network Protocols Using Snapshots}, 
  year={2022},
  volume={17},
  number={},
  pages={2673-2687},
  doi={10.1109/TIFS.2022.3192991},
  url={https://doi.org/10.1109/TIFS.2022.3192991}
}

@article{Daniele2024Survey,
author = {Daniele, Cristian and Andarzian, Seyed Behnam and Poll, Erik},
title = {Fuzzers for Stateful Systems: Survey and Research Directions},
year = {2024},
issue_date = {September 2024},
publisher = {Association for Computing Machinery},
address = {New York, NY, USA},
volume = {56},
number = {9},
issn = {0360-0300},
url = {https://doi.org/10.1145/3648468},
doi = {10.1145/3648468},
journal = {ACM Comput. Surv.},
month = {apr},
articleno = {222},
numpages = {23}
}

@ARTICLE{Fu2023HNPFuzzer,
  author={Fu, Junsong and Xiong, Shuai and Wang, Na and Ren, Ruiping and Zhou, Ang and Bhargava, Bharat K.},
  journal={IEEE Transactions on Dependable and Secure Computing}, 
  title={A Framework of High-Speed Network Protocol Fuzzing Based on Shared Memory}, 
  year={2023},
  volume={},
  number={},
  pages={1-18},
  doi={10.1109/TDSC.2023.3318571},
  url={https://doi.org/10.1109/TDSC.2023.3318571}
}

@inproceedings{Andronidis2022SnapFuzz,
author = {Andronidis, Anastasios and Cadar, Cristian},
title = {SnapFuzz: high-throughput fuzzing of network applications},
year = {2022},
isbn = {9781450393799},
publisher = {Association for Computing Machinery},
address = {New York, NY, USA},
url = {https://doi.org/10.1145/3533767.3534376},
doi = {10.1145/3533767.3534376},
booktitle = {Proceedings of the 31st ACM SIGSOFT International Symposium on Software Testing and Analysis},
pages = {340–351},
numpages = {12},
location = {Virtual, South Korea},
series = {ISSTA 2022}
}

@INPROCEEDINGS{Pham2020AFLNET,
  author={Pham, Van-Thuan and B{\"o}hme, Marcel and Roychoudhury, Abhik},
  booktitle={2020 IEEE 13th International Conference on Software Testing, Validation and Verification (ICST)}, 
  title={AFLNET: A Greybox Fuzzer for Network Protocols}, 
  year={2020},
  volume={},
  number={},
  pages={460-465},
  doi={10.1109/ICST46399.2020.00062},
  url={https://doi.org/10.1109/ICST46399.2020.00062}
}

@inproceedings{Meng2024ChatAFL,
author = {Ruijie Meng and Martin Mirchev and Marcel B{\"o}hme and Abhik Roychoudhury},
title = {Large Language Model guided Protocol Fuzzing},
booktitle = {Network and Distributed System Security (NDSS) Symposium 2024},
year = {2024},
isbn = {1-891562-93-2},
url = {https://dx.doi.org/10.14722/ndss.2024.24556},
publisher = {},
month = feb
}

@INPROCEEDINGS{Aschermann2020Ijon,
  author={Aschermann, Cornelius and Schumilo, Sergej and Abbasi, Ali and Holz, Thorsten},
  booktitle={2020 IEEE Symposium on Security and Privacy (SP)}, 
  title={Ijon: Exploring Deep State Spaces via Fuzzing}, 
  year={2020},
  volume={},
  number={},
  pages={1597-1612},
  url={https://doi.org/10.1109/SP40000.2020.00117},
  doi={10.1109/SP40000.2020.00117}
}

@InProceedings{Banks2006SNOOZE,
author="Banks, Greg
and Cova, Marco
and Felmetsger, Viktoria
and Almeroth, Kevin
and Kemmerer, Richard
and Vigna, Giovanni",
editor="Katsikas, Sokratis K.
and L{\'o}pez, Javier
and Backes, Michael
and Gritzalis, Stefanos
and Preneel, Bart",
title="SNOOZE: Toward a Stateful NetwOrk prOtocol fuzZEr",
booktitle="Information Security",
year="2006",
publisher="Springer Berlin Heidelberg",
address="Berlin, Heidelberg",
pages="343--358",
isbn="978-3-540-38343-7",
doi={10.1007/11836810\_25},
url={https://doi.org/10.1007/11836810\_25}
}

@article{Zhu2022Roadmap,
author = {Zhu, Xiaogang and Wen, Sheng and Camtepe, Seyit and Xiang, Yang},
title = {Fuzzing: A Survey for Roadmap},
year = {2022},
issue_date = {January 2022},
publisher = {Association for Computing Machinery},
address = {New York, NY, USA},
volume = {54},
number = {11s},
issn = {0360-0300},
url = {https://doi.org/10.1145/3512345},
doi = {10.1145/3512345},
journal = {ACM Comput. Surv.},
month = {sep},
articleno = {230},
numpages = {36}
}

@ARTICLE{Bohme2021Challenges,
  author={B{\"o}hme, Marcel and Cadar, Cristian and Roychoudhury, Abhik},
  journal={IEEE Software}, 
  title={Fuzzing: Challenges and Reflections}, 
  year={2021},
  volume={38},
  number={3},
  pages={79-86},
  doi={10.1109/MS.2020.3016773}
}

@INPROCEEDINGS {Schloegel2024SoK,
author = {M. Schloegel and N. Bars and N. Schiller and L. Bernhard and T. Scharnowski and A. Crump and A. Ale-Ebrahim and N. Bissantz and M. Muench and T. Holz},
booktitle = {2024 IEEE Symposium on Security and Privacy (SP)},
title = {SoK: Prudent Evaluation Practices for Fuzzing},
year = {2024},
volume = {},
issn = {2375-1207},
pages = {140-140},
doi = {10.1109/SP54263.2024.00137},
url = {https://doi.ieeecomputersociety.org/10.1109/SP54263.2024.00137},
publisher = {IEEE Computer Society},
address = {Los Alamitos, CA, USA},
month = {may}
}

@inproceedings{Gao2023Plateau,
author = {Gao, Wentao and Pham, Van-Thuan and Liu, Dongge and Chang, Oliver and Murray, Toby and Rubinstein, Benjamin I.P.},
title = {Beyond the Coverage Plateau: A Comprehensive Study of Fuzz Blockers (Registered Report)},
year = {2023},
isbn = {9798400702471},
publisher = {Association for Computing Machinery},
address = {New York, NY, USA},
url = {https://doi.org/10.1145/3605157.3605177},
doi = {10.1145/3605157.3605177},
booktitle = {Proceedings of the 2nd International Fuzzing Workshop},
pages = {47–55},
numpages = {9},
location = {Seattle, WA, USA},
series = {FUZZING 2023}
}

@inproceedings {Fioraldi2020AFL++,
author = {Andrea Fioraldi and Dominik Maier and Heiko Ei{\ss}feldt and Marc Heuse},
title = {{AFL++} : Combining Incremental Steps of Fuzzing Research},
booktitle = {14th USENIX Workshop on Offensive Technologies (WOOT 20)},
year = {2020},
url = {https://www.usenix.org/conference/woot20/presentation/fioraldi},
publisher = {USENIX Association},
month = aug
}

@inproceedings{Natella2021ProFuzzBench,
author = {Natella, Roberto and Pham, Van-Thuan},
title = {ProFuzzBench: a benchmark for stateful protocol fuzzing},
year = {2021},
isbn = {9781450384599},
publisher = {Association for Computing Machinery},
address = {New York, NY, USA},
url = {https://doi.org/10.1145/3460319.3469077},
doi = {10.1145/3460319.3469077},
booktitle = {Proceedings of the 30th ACM SIGSOFT International Symposium on Software Testing and Analysis},
pages = {662–665},
numpages = {4},
location = {Virtual, Denmark},
series = {ISSTA 2021}
}

@inproceedings{Schumilo2022Nyx-net,
author = {Schumilo, Sergej and Aschermann, Cornelius and Jemmett, Andrea and Abbasi, Ali and Holz, Thorsten},
title = {Nyx-net: network fuzzing with incremental snapshots},
year = {2022},
isbn = {9781450391627},
publisher = {Association for Computing Machinery},
address = {New York, NY, USA},
url = {https://doi.org/10.1145/3492321.3519591},
doi = {10.1145/3492321.3519591},
booktitle = {Proceedings of the Seventeenth European Conference on Computer Systems},
pages = {166–180},
numpages = {15},
location = {Rennes, France},
series = {EuroSys '22}
}

@inproceedings{Bohme2022benchmarking,
author = {B\"{o}hme, Marcel and Szekeres, L\'{a}szl\'{o} and Metzman, Jonathan},
title = {On the reliability of coverage-based fuzzer benchmarking},
year = {2022},
isbn = {9781450392211},
publisher = {Association for Computing Machinery},
address = {New York, NY, USA},
url = {https://doi.org/10.1145/3510003.3510230},
doi = {10.1145/3510003.3510230},
booktitle = {Proceedings of the 44th International Conference on Software Engineering},
pages = {1621–1633},
numpages = {13},
location = {Pittsburgh, Pennsylvania},
series = {ICSE '22}
}

@article{Miller1990Fuzzing,
author = {Miller, Barton P. and Fredriksen, Lars and So, Bryan},
title = {An empirical study of the reliability of UNIX utilities},
year = {1990},
issue_date = {Dec. 1990},
publisher = {Association for Computing Machinery},
address = {New York, NY, USA},
volume = {33},
number = {12},
issn = {0001-0782},
url = {https://doi.org/10.1145/96267.96279},
doi = {10.1145/96267.96279},
journal = {Commun. ACM},
month = {dec},
pages = {32–44},
numpages = {13}
}

@article{brown2020language,
  title={Language models are few-shot learners},
  author={Brown, Tom and Mann, Benjamin and Ryder, Nick and Subbiah, Melanie and Kaplan, Jared D and Dhariwal, Prafulla and Neelakantan, Arvind and Shyam, Pranav and Sastry, Girish and Askell, Amanda and others},
  journal={Advances in neural information processing systems},
  volume={33},
  pages={1877--1901},
  year={2020}
}

@Article{Zhang2023Survey,
AUTHOR = {Zhang, Zhaowei and Zhang, Hongzheng and Zhao, Jinjing and Yin, Yanfei},
TITLE = {A Survey on the Development of Network Protocol Fuzzing Techniques},
JOURNAL = {Electronics},
VOLUME = {12},
YEAR = {2023},
NUMBER = {13},
ARTICLE-NUMBER = {2904},
URL = {https://www.mdpi.com/2079-9292/12/13/2904},
ISSN = {2079-9292},
DOI = {10.3390/electronics12132904}
}

@misc{AFL,
  author = {Google},
  title = {AFL - American Fuzzy Lop},
  year = {2024},
  howpublished = {\url{https://github.com/google/AFL}},
  note = {Accessed: 2024-07-15}
}

@article{chen2018systematic,
  title={A systematic review of fuzzing techniques},
  author={Chen, Chen and Cui, Baojiang and Ma, Jinxin and Wu, Runpu and Guo, Jianchao and Liu, Wenqian},
  journal={Computers \& Security},
  volume={75},
  pages={118--137},
  year={2018},
  publisher={Elsevier}
}

@article{chipounov2011s2e,
  title={S2E: A platform for in-vivo multi-path analysis of software systems},
  author={Chipounov, Vitaly and Kuznetsov, Volodymyr and Candea, George},
  journal={Acm Sigplan Notices},
  volume={46},
  number={3},
  pages={265--278},
  year={2011},
  publisher={ACM New York, NY, USA}
}

@article{nethercote2007valgrind,
  title={Valgrind: a framework for heavyweight dynamic binary instrumentation},
  author={Nethercote, Nicholas and Seward, Julian},
  journal={ACM Sigplan notices},
  volume={42},
  number={6},
  pages={89--100},
  year={2007},
  publisher={ACM New York, NY, USA}
}

@inproceedings{kaksonen2001software,
  title={Software security assessment through specification mutations and fault injection},
  author={Kaksonen, Rauli and Laakso, Marko and Takanen, Ari},
  booktitle={Communications and Multimedia Security Issues of the New Century: IFIP TC6/TC11 Fifth Joint Working Conference on Communications and Multimedia Security (CMS’01) May 21--22, 2001, Darmstadt, Germany},
  pages={173--183},
  year={2001},
  organization={Springer}
}

@misc{boofuzz,
author = {Joshua Pereyda},
title = {boofuzz Documentation},
howpublished={\url{https://github.com/jtpereyda/boofuzz}},
year = {2019}
}

@ARTICLE{Pham2021Smart,
  author={Pham, Van-Thuan and Böhme, Marcel and Santosa, Andrew E. and Căciulescu, Alexandru Răzvan and Roychoudhury, Abhik},
  journal={IEEE Transactions on Software Engineering}, 
  title={Smart Greybox Fuzzing}, 
  year={2021},
  volume={47},
  number={9},
  pages={1980-1997},
  doi={10.1109/TSE.2019.2941681}}

@inproceedings{serebryany2012addresssanitizer,
  title={$\{$AddressSanitizer$\}$: A fast address sanity checker},
  author={Serebryany, Konstantin and Bruening, Derek and Potapenko, Alexander and Vyukov, Dmitriy},
  booktitle={2012 USENIX annual technical conference (USENIX ATC 12)},
  pages={309--318},
  year={2012}
}

\end{document}